\documentclass[twocol]{ametsocV6.1}

\usepackage{multicol}

\title{Mid-latitude interactions expand the Hadley circulation}

\authors{W. Moon,\aff{a} J. S. Wettlaufer,\aff{b,c}\correspondingauthor{John S. Wettlaufer, jw@fysik.su.se, john.wettlaufer@yale.edu} 
}

\affiliation{\aff{a}{Department of Environmental Atmospheric Sciences, Pukyong National University, 48513 Busan, South Korea}\\
\aff{b}{Yale University, 210 Whitney Avenue, New Haven, USA}\\
\aff{c}{Nordita, Stockholm University, Hannes Alfv{\'e}ns v{\"a}g 12, 114 19 Stockholm, Sweden}\\
}

\abstract{The Hadley circulation describes a planetary-scale tropical atmospheric flow, which has a major influence on climate.  
Contemporary theoretical understanding is based upon angular momentum conservation, the basic dynamical constraint governing the state of the flow pattern,
and scaling relationships characterizing the macroturbulence created by synoptic eddies.
However, despite the degree of success in representing the Hadley circulation, the canonical theoretical model does not treat interactions with other regions, 
particularly the mid-latitudes. Here, we extend the original model of \citet{held1980}
 to include the influence of mid-latitude large-scale atmospheric dynamics, which we treat using the planetary-scale heat equation 
 with a parameterized poleward heat flux driven by synoptic eddies. The energy flux balance within the Hadley cell
 includes the poleward heat flux at the poleward edge of the cell, which is controlled by the baroclinic instability of the sub-tropical jet.   
 We find that an increase (decrease) in the poleward heat flux leads to a strengthening (weakening) tropical convection, driving an equatorward (poleward) shift of the edge of the Hadley cell.
 Thus, our theoretical solutions suggest that global warming, which can reduce the baroclinicity of the subtropical jet, 
 can lead to the poleward expansion of the Hadley cell due to the change in energy flux balance within it.  } 

\begin{document}

%% Necessary!
\maketitle

%%%%%%%%%%%%%%%%%%%%%%%%%%%%%%%%%%%%%%%%%%%%%%%%%%%%%%%%%%%%%%%%%%%%%
% SIGNIFICANCE STATEMENT/CAPSULE SUMMARY
%%%%%%%%%%%%%%%%%%%%%%%%%%%%%%%%%%%%%%%%%%%%%%%%%%%%%%%%%%%%%%%%%%%%%
%
% If you are including an optional significance statement for a journal article or a required capsule summary for BAMS 
% (see www.ametsoc.org/ams/index.cfm/publications/authors/journal-and-bams-authors/formatting-and-manuscript-components for details), 
% please apply the necessary command as shown below:
%
% Significance Statement (all journals except BAMS)
%
\statement
The Hadley circulation is a principal dynamical feature in the tropics, acting as an engine that transfers heat and momentum
to high latitudes. 
At the same time, the extratropical regions can influence the Hadley circulation.  Under global warming, high latitudes warm faster than low latitudes, reducing the equator to pole temperature difference and the poleward heat flux in the mid-latitudes.   The theory described here shows that the decrease in the poleward heat flux can lead to the expansion of the Hadley cell, and a poleward shift of subtropical dry zones.
%
%% Capsule (BAMS only)
%%
%\capsule
%       Enter BAMS capsule here, no more than 30 words. See \url{www.ametsoc.org/index.cfm/ams/publications/author-information/formatting-and-manuscript-components/#capsule} for details.
% 
%% * * If using twocol mode, you will need to use the commands "twocolsig" and "twocolcapsule" in place of "sig" and "capsule"
%%      to ensure that the text box correctly spans across both columns.

%%%%%%%%%%%%%%%%%%%%%%%%%%%%%%%%%%%%%%%%%%%%%%%%%%%%%%%%%%%%%%%%%%%%%
% MAIN BODY OF PAPER
%%%%%%%%%%%%%%%%%%%%%%%%%%%%%%%%%%%%%%%%%%%%%%%%%%%%%%%%%%%%%%%%%%%%%
%
\section{Introduction \label{sec:introduction}}
The imbalance of energy between the tropics and the high-latitudes drives large-scale atmospheric and oceanic motions 
under the influence of Earth's rotation \citep[e.g.,][]{trenberth1994}.  These motions can be described by solving the primitive equations,
which are approximate forms of the Navier-Stokes equations and the conservation of mass and thermal energy in a rotating frame of reference, assuming horizontal motions dominate vertical motions for an atmosphere of depth much less than Earth's radius \citep{phillips1959}.
Their numerical solution was first achieved by the revolutionary development of stored program computing \citep[e.g.,][]{edwards2011}. 
However, even with this computational advance, the complexity of the primitive equations still masks a great deal of understanding.  
This motivates theoretical approaches that model the essential physical processes \citep{houghton1997}. 

The zonal mean dynamics of the tropical atmosphere is described by a thermally-driven cell extending from the equator to the sub-tropics \citep[e.g.,][]{webster2004}. 
Tropical convection drives upward motions with the upper atmosphere traveling towards high-latitudes, 
where radiative cooling drives downward motions in sub-tropical regions. The air returns to the tropics through the near surface trade winds, 
thereby completing what is known as the Hadley circulation. The simplest model, due to \citet{north1975}, is a forced heat equation 
with a thermal eddy diffusivity $D$ that parameterizes the activity of synoptic eddies.
The poleward heat transport by {\color{black}synoptic eddies} is interpreted as being ``diffusive'' \citep[e.g.,][]{okubo2001}, 
under the assumption that turbulent eddies govern the equator-to-pole transport.  
However, turbulent diffusion cannot characterize the entire transport process of the large-scale convective cell 
extending from the tropics to the mid-latitudes.  

The central means of side-stepping the complexity and ambiguities of turbulent diffusivity in the Hadley circulation is 
to replace the equations of motion themselves with angular momentum conservation.  If $a$ and $\Omega$ are Earth's radius and rotation rate respectively, 
then a ``static air parcel'' (i.e., no zonal mean zonal wind) at the surface in the tropics has angular momentum $\Omega a^2$, which is conserved during its journey towards the upper edge of the Hadley cell.  
This is used to calculate the zonal mean structure of the Hadley circulation \citep{held1980}. Combined with energy flux conservation inside the cell, the theory of \citet{held1980} 
deduces the size of the Hadley cell in terms of the position of the descending edge.  This idealized model was extended by \citet{lindzen1988} 
to represent the seasonal evolution of the Hadley circulation. 
{\color{black} Related analytical work includes, for example, the viscous solution of the Charney problem (i.e., symmetric circulation) for small Rossby number \citep{fang1994}, and single-layer axisymmetric \citep{sobel2009}, and shallow water \citep{adam2013} models.}

%{\color{red} Later, there has been several research to seek analytic solutions based on reduced equations \citep{fang1994}, a shallow water model \citep{adam2013}, or a single-layer axisymmetric model \citep{sobel2009}.}

Recently \citet{hill2022} used a different approach to deduce the size of the Hadley cell.  Numerical studies 
showed that angular momentum is not conserved in the upper branch of the Hadley cell \citep{walker2006}, 
which was expected due to the dynamic role of synoptic eddies leading to the slowdown of the zonal mean zonal wind.
Hence, rather than relying on angular momentum conservation combined with thermodynamic constraints, the descending edge
is determined as the minimum latitude at which the zonal mean wind is baroclinically unstable \citep{kang2012}, and
the ascending edge is determined by calculating the latitude at which the absolute vorticity vanishes near the tropopause \citep{hill2021}.   

The dynamics of baroclinic waves in the mid-latitudes are known to influence the intensity and extent of the Hadley cell \citep{ lee2003, walker2006}.  
{\color{black}Baroclinic wave activity, represented by baroclinicity or supercriticality in subtropical regions, is argued to be strongly associated with the state of the Hadley circulation \citep{korty2008,levine2015},
and the eddy fluxes induced by unstable baroclinic waves play an important role in the characteristics of the Hadley cell \citep{singh2016, singh2017}. 
}

%{\color{red}Baroclinic wave activity represented by baroclinicity or supercriticality in subtropical areas seems 
%to be strongly associated with the status of the Hadley circulation \citep{korty2008,levine2015}
%and the eddy fluxes induced by unstable baroclinic waves play an important role on the Hadley cell \citep{singh2016, singh2017}. 
%}
Indeed, the dynamic and thermodynamic interaction between the tropics and mid-latitudes is a key factor in determining both the large scale structure 
and internal dynamics of the Hadley cell. Moreover, observations and global climate models show that the extent of the Hadley cell has increased 
due to global warming \citep{frierson2007a, lu2007, johanson2009, choi2014}. This can lead to a dramatic change in large-scale tropical atmospheric dynamics, 
including the monsoon and impacts on the desertification in sub-tropical regions. The Hadley circulation is not a closed subsystem of  the climate, 
and given its importance, an appropriate theory should contain dynamic and thermodynamic interactions between the
 tropics and the mid-latitudes.
 
 Here, we extend the closed Hadley circulation model to include the interaction with mid-latitudes. 
 This influence is approximated by the planetary-scale heat equation with a parameterization
 of the synoptic-scale turbulent heat flux. The mutual interaction between the Hadley cell and the mid-latitude atmosphere is considered through two thermodynamic constraints; (a) the continuity of potential temperature at the poleward edge of the Hadley cell, and (b) the energy flux balance inside the cell, including the poleward heat flux calculated at the edge of the cell.  
Namely, the deviation of the potential temperature at the cell edge from radiative-convective equilibrium drives the large-scale dynamics of {\color{black}synoptic eddies} in the mid-latitudes, and the poleward heat flux from synoptic eddies balances the energy flux inside the cell.  Therefore, these mid-latitude interactions are viewed as an extension of the theory of \citet{held1980}, 
 and a modification of the energy flux balance model of \citet{north1975}.   
 
 {\color{black} 
 The planetary-scale heat equation plays a critical role in introducing eddy diffusivities, which approximate turbulent heat transfer by synoptic eddies. Employing eddy diffusion as a parameterization for turbulent heat transfer requires a clear scale separation between the mean field and the eddies. Specifically, the temporal and spatial scales of the mean field must be significantly larger than those of the eddies to ensure that the mean field remains relatively unchanged while it accumulates the effects of synoptic eddies.

Consequently, the equations that govern the Hadley circulation and the mid-latitudes must operate on larger temporal and spatial scales than the quasi-geostrophic equations, which are designed to capture synoptic eddy dynamics. Previous studies have established a theoretical foundation for this approach using a multi-scale formalism, which supports the use of planetary geostrophic motion as a mean field for synoptic-scale processes \citep{dolaptchiev2013,moon2020}.

Building on this formalism, the quasi-oscillatory behavior of anomalous potential temperature in the mid-latitudes has been explained through the planetary-scale heat equation, incorporating parameterized eddy heat fluxes and memory effects \citep{moon2021}. This framework underscores the suitability of planetary geostrophic motion for investigating the interactions between the Hadley cell and the mid-latitudes, particularly in relation to turbulent heat transfer by synoptic eddies near the cell edge.
 }
 
 {\color{black}In order to make this paper reasonably self-contained and accessible to readers who are not intimate with these studies, in section \ref{sec:model} we provide general background on the geophysical fluid dynamics and multi-scale analysis necessary to understand our main contribution, which is given in section \ref{sec:HHPG}.  Readers intimate with the \citet{held1980} theory, planetary- and quasi-geostrophic motion, large-scale heat transport, and multi-scale asymptotic analysis may wish to begin with section \ref{sec:HHPG}.}

 \section{Theoretical background\label{sec:model}}
 
 The Hadley circulation emerges from the interaction between {\color{black}synoptic eddies} and tropical convection. 
 Despite this complexity, a key dynamic constraint, angular momentum conservation, facilitates the deduction of the meridional structure of potential temperature.
The foundational theory of \cite{held1980} treats the tropical regions adjacent to the mid-latitudes, maintaining a radiative-convective equilibrium temperature, 
defined as $\Theta_E$. This equilibrium temperature is determined through a regional energy flux balance, 
accounting for local convection with a length-scale asymptotically smaller than the synoptic scale. Notably, the original theory 
does not incorporate the potential impact of synoptic eddies, generated primarily by baroclinic instability near the edge of the Hadley circulation.
Studies highlight the influence of synoptic eddy dynamics on the dynamical characteristics of the Hadley circulation \citep{ lee2003, walker2006}. 
Therefore, it is important to consider the effect of mid-latitude synoptic eddies on the overall dynamics of the Hadley circulation.
 
 Atmospheric dynamics on the planetary scale, characterized by a length-scale larger than the external Rossby deformation radius (3,000 km), 
 is governed by planetary geostrophic motion, a heat equation and hydrostatic balance \citep[e.g.,][and refs therein]{moon2020}.
{\color{black} The Rossby deformation radius represents the length scale at which rotational dynamics become as significant as gravitational forces. 
In particular, the external Rossby deformation radius is derived under the assumption that the large-scale atmosphere is barotropic.}
 In contrast, synoptic-scale atmospheric motions evolve according to the potential vorticity equation \citep{burger1958, phillips1963}.
Given the size of the Hadley circulation, falling within the planetary scale, the leading-order dynamics should be represented
by planetary geostrophic motion, {\color{black}{although in the tropics that balance is disrupted because the Coriolis effect becomes negligible, which underlies the challenges of expressing the dynamical interaction between the tropics and the extra-tropics.  Moreover, planetary geostrophic motion alone cannot comprehensively capture the dynamics of the Hadley circulation, 
because it does not account for synoptic eddies}}.
Despite the overall planetary-scale structure, the synoptic dynamics play a crucial role in shaping the major characteristics of the circulation. 
Advancements employing a multi-scale formalism, either focus solely on the tropics \citep{Majda2003}, or propose that planetary geostrophic motion, coupled with thermal forcing, 
establishes a balanced field conducive to the growth of synoptic eddies \citep{dolaptchiev2013,moon2020}. 
This latter multi-scale formalism provides a framework for describing wave-mean flow interactions, 
suggesting that the Hadley circulation is a consequence of such interactions.
In this section, we elaborate on the original \cite{held1980} theory within the context of planetary-scale dynamics, and extend the treatment to include the influence of synoptic eddies in the mid-latitudes.
%incorporating angular momentum conservation. Furthermore, we extend the treatment to include the influence of synoptic eddies in the mid-latitudes.

We now summarize the essential features here.

 \subsection{Planetary geostrophic motion and its interaction with synoptic eddies--motivation}
 
The theory of large-scale atmospheric dynamics is rooted in the principles of Rossby wave dynamics \citep{dickinson1978}. 
Disturbances acting upon a mean flow field, characterized by a background potential vorticity gradient, give rise to large-scale waves 
that exert control over synoptic mid-latitude weather patterns. The generation, propagation, and breaking of these waves significantly 
influence the overall circulation of the large-scale atmosphere. The dynamics of Rossby waves are effectively described by the quasi-geostrophic potential vorticity equation, 
which provides a fundamental framework for investigating various large-scale phenomena in the atmosphere, 
primarily due to the pivotal role played by large-scale eddies \citep{pedlosky2013}.
  
However, the understanding of the general circulation, including the Hadley cell, entails recognizing it 
as an organized structure arising from the collective influence of turbulent eddies. 
To effectively represent the comprehensive structure of the general circulation, 
it is essential to establish a framework capable of capturing the cumulative impact of {\color{black}synoptic eddies}.   This requires a mathematical treatment 
with spatial and temporal scales that are asymptotically larger than the synoptic scale, thereby parameterizing the cumulative effect of turbulent eddies through the gradient of the state variables.

In the context of the Fickian approximation, achieving scale separation is crucial. Thus, planetary geostrophic motion, 
characterized by an asymptotically larger length scale than synoptic-scale eddies, provides a suitable setting
for applying the Fickian approximation to the turbulent heat flux induced by eddies.

\subsection{Quasi-geostrophic motion, Planetary geostrophic motion, and the heat equation}

Quasi-geostrophic motion and planetary geostrophic motion are characterized by a small Rossby number, so that the Coriolis effect dominates inertial effects. 
Thus, both scales exhibit the same leading-order balances in momentum conservation; geostrophic and hydrostatic balances.  

In quasi-geostrophic motion, the leading-order balance does not explicitly involve the continuity and heat equations, which are automatically satisfied by geostrophic balance. 
Therefore, the leading-order balance alone cannot uniquely determine the leading-order variables, which are the geostrophic winds and pressure, a unique determination of which requires 
higher-order dynamics, through the conservation of potential vorticity.

On the other hand, when the horizontal length-scale is comparable to the external Rossby deformation radius $L_D \equiv \sqrt{gH}/f$, where $H$ is a characteristic vertical scale, and $f$ is the mid-latitude Coriolis parameter, the continuity and heat equations are not satisfied automatically by geostrophic balance.  Thus, using a linear meridional variation in the Coriolis parameter, $\beta \equiv df/dy$, in the continuity equation has a leading order effect, which leads to the Sverdrup relationship \citep{sverdrup1947}.   Note that $L_D \approx 3000$ km, and characterizes the length scale of a Rossby wave for a barotropic atmosphere on Earth.  

Importantly, we turn to the heat equation to show a clear separation of scales.  Namely, whereas on the quasi-geostrophic scale the advection of potential temperature appears as part of a higher-order balance, on the planetary-scale the advection of potential temperature appears at leading-order.  Therefore, the heat equation is a central focus in planetary geostrophic motion.

We non-dimensionalize the primitive equations with length-scale $L$, velocity-scale $U$, and advective time-scale $L/U$, where the dimensional independent variables $x^*,y^*,z^*$ and $t^*$ have their usual meanings as zonal, meriodional, vertical and time coordinates respectively. 
The dimensional pressure is $P^*=P^*(x^*,y^*,z^*,t^*)$ and the dimensionless pressure is $p=p(x, y, z, t)$, and they are related through $P^*= P^*_S(z^*)+\rho^*_S(z^*)Uf_0Lp$, where $P^*_S(z^*)$ and $\rho^*_S(z^*)$ are horizontally averaged (the subscript $S$) reference 
vertical profiles of pressure and density respectively,  $f_0$ is the Coriolis parameter at a reference latitude, and thus
$\rho^*_S(z^*)Uf_0L$ is a scale for the anomalous pressure originating from geostrophic balance.  
The dimensional density, $\rho^*$, and potential temperature, $\Theta^*$, are non-dimensionalized using the geostrophic and hydrostatic balances, giving
$\rho^*=\rho^*_S(z^*)[1+\epsilon F\rho]$ and $\Theta^*=\Theta^*_S(z^*)[1+\epsilon F\Theta]$, where $\rho$ and $\Theta$ are dimensionless.
Here, the Rossby number is $\epsilon \equiv U/fL$, and the Froude number is $F \equiv f^2L^2/gH=L^2/L_D^2$,
with $g$ the acceleration due to gravity.  Based on this scaling the primitive equations become
\begin{align}
&\epsilon \frac{Du}{Dt}-(1+\beta y)v=-\frac{\partial p}{\partial x}, \nonumber \\
&\epsilon \frac{Dv}{Dt}+(1+\beta y)u=-\frac{\partial p}{\partial y}, \nonumber \\
& \frac{\partial u}{\partial x}+\frac{\partial v}{\partial y}+\frac{1}{\rho_S}\frac{\partial}{\partial z}(\rho_Sw)=0, \nonumber \\
&\rho = \frac{1}{\rho_S}\frac{\partial}{\partial z}(\rho_S p), \qquad \text{and} \nonumber \\
& \epsilon F \frac{D\Theta}{Dt}+\frac{1}{\Theta_S}\frac{d\Theta_S}{dz}w = Q,
\label{eq:prim}
\end{align}
 where $\vec{u}=(u,v,w)$ is the velocity field,  $\frac{D}{Dt} \equiv \frac{\partial}{\partial t}+u\frac{\partial}{\partial x}+v\frac{\partial}{\partial y}+w\frac{\partial}{\partial z}$, 
 and $Q=Q(x, y, z, t)$ is the dimensionless equator to pole radiative-convective energy flux, which drives the meridional temperature gradient, and is $O(\epsilon^2)$.
 A more detailed development can be found in \citet{pedlosky2013}.
  
 As noted above, large-scale atmospheric dynamics is characterized by a small Rossby number $\epsilon \ll 1$, which is the basis for a regular perturbation analysis, expanding the dynamic and thermodynamic variables as $x=x_0+\epsilon x_1 + O(\epsilon^2)$, yielding the leading-order behavior of geostrophic and hydrostatic balance as
\begin{align}
v_0 = \frac{\partial p_0}{\partial x}, \quad u_0=-\frac{\partial p_0}{\partial y}, ~~ \text{and} ~~ \rho_0 = \frac{1}{\rho_S}\frac{\partial}{\partial z}(\rho_S p_0).
\end{align} 
Next, we seek to extract a leading-order balance from the heat equation, which is determined by the choice of $L$.  
 
 \subsubsection{Length scale selection and the heat equation}
 
Selecting $L$ as the internal Rossby deformation radius, which is approximately $1000$ km, yields $F$ of approximately $0.1$, which is also the same order of magnitude as $\epsilon$, and  the static stability term $\frac{1}{\Theta_S}\frac{d\Theta_S}{dz}$.  Therefore, the temperature advection term is $\epsilon F \frac{D\Theta}{Dt} = O(\epsilon^2)$, so that under adiabatic conditions the leading-order vertical velocity $w_0$ is essentially zero, which characterizes quasi-geostrophic motion.

Quasi-geostrophic motion is based on leading order hydrostatic and geostrophic balances.  However, for a given set of boundary conditions,  this cannot fully determine all of the fields described by Eqs. \eqref{eq:prim}, resulting in so-called geostrophic degeneracy. 
Hence, in order to determine the leading-order variables, one must consider dynamics at  $O(\epsilon)$.  This leads to quasi-geostrophic potential vorticity conservation under adiabatic conditions, 
which captures the essentials of quasi-geostrophic motion.

Alternatively, selecting $L$ to be of order the external Rossby deformation radius, $L_D$, gives $F$ = $O(1)$. This leads to $\epsilon F \frac{D\Theta}{Dt}$ and $\frac{w}{\Theta_S}\frac{d\Theta_S}{dz}$ being the same order of magnitude.  Therefore, on this length-scale, geostrophic degeneracy vanishes, thereby eliminating the need to consider higher-order dynamics \citep{pedlosky2013, moon2020}.

\subsubsection{Planetary geostrophic motion--multi-scale analysis\label{sec:PG}}

Planetary geostrophic motion is comprised of geostrophic and hydrostatic balance, 
the Sverdrup relation arising from continuity, and the heat equation, and is given in non-dimensional form as follows:
 \begin{align}
  &u_L = -\frac{\partial P_L}{\partial y}, \quad v_L = \frac{\partial P_L}{\partial x}, \quad \rho_L = -\frac{1}{\rho_S}\frac{\partial}{\partial z}(\rho_SP_L), \nonumber \\
  &\frac{1}{\rho_S}\frac{\partial}{\partial z}(\rho_S w_L) - \beta_L v_L = 0, \qquad \text{and} \qquad \nonumber \\
  &\frac{\partial\Theta_L}{\partial t} + u_L\frac{\partial\Theta_L}{\partial x} + v_L\frac{\partial\Theta_L}{\partial y} + w_L\left(S(z) + \frac{\partial\Theta_L}{\partial z}\right) = Q_L,
\end{align}
where the subscripts $L$ denote planetary-scale.  Thus, the pressure is $P_L$, the potential temperature is $\Theta_L = \frac{\partial P_L}{\partial z}$,  the zonal, meridional, and vertical velocities are $u_L$, $v_L$, and $w_L$ respectively, and the planetary-scale thermal forcing is $Q_L$.  The $\beta$-effect is characterized by $\beta_L$, the hemispheric average vertical density profile is $\rho_S$, 
and the dry static stability of the entire hemisphere is given by  $S(z)$.

Central here is the heat equation for $\Theta_L$ under the influence of the velocity field, and the thermal forcing $Q_L$. Therefore, planetary geostrophic motion yields a balanced flow in response to a specified planetary-scale thermal forcing  \citep{pedlosky2013, moon2020}.

A complete description of the dynamics of the large-scale atmosphere must treat the interaction between the planetary and synoptic scales, which requires a multi-scale formalism,  
the crux of which lies in expressing thermodynamic variables $P$, $\Theta$, and $\rho$ as a linear sum of planetary and synoptic scales \citep{moon2020}.
For instance, $P^*=P^*_S(z^*)+\rho^*_S(z^*)f_0L_DUP_L(X,Y,z,\tilde{t})+\rho^*_S(z^*)f_0LUP(x,y,z,t)$, where as above $L_D$ is the external Rossby deformation radius ($\approx 3000 \, \text{km}$), 
and $L$ is the length for the synoptic scale ($\approx 1000 \, \text{km}$). The symbols $X$, $Y$, and $\tilde{t}$ ($x$, $y$, and $t$) denote the
space and time variables for planetary-scale (synoptic-scale) motions. The multi-scale formalism leads to
\begin{align}
  &u_L = -\frac{\partial P_L}{\partial Y}, \quad v_L = \frac{\partial P_L}{\partial X}, \quad u_0 = -\frac{\partial p_0}{\partial y}, \quad \text{and~~} v_0 = \frac{\partial p_0}{\partial x}, \nonumber \\
  &\frac{1}{\rho_S}\frac{\partial}{\partial z}(\rho_Sw_L) = \beta_Lv_L, \nonumber \\
  &\frac{\partial\Theta_L}{\partial\tilde{t}}+u_L\frac{\partial\Theta_L}{\partial X}+v_L\frac{\partial\Theta_L}{\partial Y}+w_L\left(\frac{\partial\Theta_L}{\partial z}+S(z)\right) \nonumber \\
  &=-\left(\frac{\partial}{\partial t}+u_L\frac{\partial}{\partial x}+v_L\frac{\partial}{\partial y}\right)\Theta_0
  -\left(u_0\frac{\partial}{\partial X}+v_0\frac{\partial}{\partial Y}\right)\Theta_L \nonumber \\
  &-\frac{\partial}{\partial x}(u_0\Theta_0)-\frac{\partial}{\partial y}(v_0\Theta_0)-w_1\left(\frac{\partial\Theta_L}{\partial z}+S(z)\right)+Q_L, \quad \text{and~~}  \nonumber \\
  &\frac{\partial}{\partial t}\nabla^2P_0+(u_L+u_0)\frac{\partial}{\partial x}\nabla^2P_0+(v_L+v_0)\frac{\partial}{\partial y}\nabla^2P_0 \nonumber \\
  &+\beta\frac{\partial P_0}{\partial x} = \frac{1}{\rho_S}\frac{\partial}{\partial z}(\rho_S w_1).
\end{align}
Here, the subscripts $0$ and $1$ denote the leading order and $O(\epsilon)$ synoptic scale variables, respectively.  
The planetary scale continuity equation transforms into the Sverdrup relation.  Both scales obey geostrophic and hydrostatic balances, and both scales collectively contribute to the heat flux balance. 
In terms of planetary-scale motion, synoptic-scale motions act as external forcing for the spatio-temporal evolution of planetary-scale potential temperature. 
Finally, planetary-scale motion acts as a mean field for the horizontal vorticity of synoptic-scale flows \citep{moon2020}.

The general circulation in the atmosphere is treated using zonally-averaged fields on the hemispheric spatial scale. In particular, the Hadley circulation is a consequence of tropical convection and {\color{black}synoptic eddies} in mid-latitudes.  Since the formulation is based on the beta plane, it is challenging to fully account for the influence of tropical convection. 
The \citet{held1980} theory employs angular momentum conservation to establish a connection between tropical convection 
and the dynamics of mid-latitudes. Therefore, the dynamics of the Hadley circulation can be effectively described by combining 
angular momentum conservation, planetary geostrophic motion, and synoptic eddies.

We treat the effect of synoptic eddies in the framework of planetary geostrophic motion by taking the spatio-temporal and zonal averages of synoptic variables, which leads to
\begin{align}
    &u_L = -\frac{\partial P_L}{\partial Y}, \nonumber \\
    &\Theta_L = \frac{\partial P_L}{\partial z}, \nonumber \\
    &\frac{1}{\rho_S}\frac{\partial}{\partial z}(\rho_S\langle\overline{w}_1\rangle) = -\epsilon\frac{\partial^2}{\partial Y^2}\langle\overline{u_0v_0}\rangle, \qquad \text{and} \nonumber \\
    &\frac{\partial\Theta_L}{\partial\tilde{t}}=Q_L-\epsilon^{1/2}\frac{\partial}{\partial Y}\langle\overline{v_0\Theta_0}\rangle
       -\langle\overline{w}_1\rangle\left(\frac{\partial\Theta_L}{\partial z}+S(z)\right),
       \label{eq:full}
\end{align}
where $\langle X\rangle$ and $\overline{X}$ denote zonal and time averages respectively, and $\epsilon$ is the planetary scale Rossby number. It appears from the spatial average of the synoptic spatial derivative $\partial/\partial y$, which becomes $\epsilon^{1/2}\partial/\partial Y$ wherein $L/L_D = \epsilon^{1/2}$.
This treatment consists of geostrophic balance containing the westerlies in the mid-latitudes, and heat flux balance, which has three components: thermal forcing $Q_L$, poleward heat flux by synoptic eddies, and adiabatic warming or cooling by vertical motion. This vertical motion is driven by eddy momentum convergence, which is equivalent to a horizontal vorticity flux.

Within this framework the general circulation of the large-scale atmosphere can be understood by examining how the thermal forcing, $Q_L$, is 
distributed by {\color{black}synoptic eddies}. In particular, the Hadley circulation can be described by combining the above framework with angular momentum conservation.

\subsubsection{Radiative convective equilibrium temperature\label{sec:RC}} 

Within the planetary geostrophic framework, we seek to distinguish the roles played by thermodynamics and fluid dynamics 
in the distribution of the thermal forcing, specifically in the form of shortwave radiance. Within the framework of quasi-geostrophic dynamics, 
the role of radiative-convective equilibrium is indirectly characterized as thermal forcing. The potential temperature exhibits a tendency 
to relax towards the radiative-convective equilibrium temperature with a characteristic response time-scale $\tau$ \citep{held1994}.
This process can be systematically derived within the context of planetary geostrophic motion, 
as this motion is directly influenced by the residual effects of local radiation and convection.

The radiative-convective temperature, $\Theta_E$ ,  is a consequence of the local balance between radiative heat flux and convection. 
The actual temperature differs from the radiative-convective temperature due to the presence of large-scale atmospheric flow. 
The primary function of this large-scale atmospheric flow is to transfer surplus energy from low-latitudes to high-latitudes. 
Consequently, one anticipates that the real temperature is lower (higher) than the radiative-convective equilibrium temperature in the low-latitudes (high-latitudes).
However, the strength of the seasonal cycle in surface temperature indicates that the dominant factor influencing the local temperature 
is the local energy flux balance, primarily determined by shortwave radiation. As a result, the radiative-convective equilibrium temperature
is much greater than the difference between the real temperature $\Theta_L$ and $\Theta_E$.

In the heat equation, the external thermal forcing $Q_L$ represents local radiative-convective processes. 
It is reasonable to assume that $Q_L = Q_L(\Theta_L)$, so that the definition of radiative-convective equilibrium is $Q_L(\Theta_E) = 0$. 
Therefore, $Q_L(\Theta_L) \simeq Q_L(\Theta_E)  + \partial Q_L/\partial\Theta_L|_{\Theta_L=\Theta_E} (\Theta_L - \Theta_E) = -(\Theta_L - \Theta_E)/\tau$, which is
the sensitivity of the local radiative-convective equilibrium, with response time-scale of $\tau$.

{\color{black} We define the anomalous potential temperature as $\eta_L \equiv \Theta_L - \Theta_E$, and given the dominance of the seasonal cycle, we let $\Theta_E \gg \eta_L$, so that $\Theta_L \simeq \Theta_E+\sqrt{\epsilon}\eta_L$, and thus the heat equation from \eqref{eq:full} becomes}
\begin{align}\label{eq:ano_heat}
    \frac{\partial\eta_L}{\partial\tilde{t}} = -\frac{1}{\tau}\eta_L - \frac{\partial}{\partial Y}\langle\overline{v_0\Theta_0}\rangle
    - \epsilon^{-1/2}\langle\overline{w}_1\rangle\left(\frac{\partial\Theta_E}{\partial z} + S(z)\right),
\end{align}
where $\Theta_L \simeq \Theta_E + \epsilon^{1/2}\eta_L$, so that $\frac{\partial\Theta_L}{\partial\tilde{t}}\simeq \epsilon^{1/2} \frac{\partial\eta_L}{\partial\tilde{t}}$ and $Q_L(\Theta_L) \simeq  -\frac{\epsilon^{1/2}}{\tau}\eta_L$.  Note that all terms are $O(\epsilon)$, apart from the last term proportional to $\langle\overline{w}_1\rangle$, which is 
 $O(\epsilon^{1/2})$ because $\langle\overline{w}_1\rangle \sim O(\epsilon)$, in accordance with the relationship 
between $\langle\overline{w}_1\rangle$ and the eddy momentum flux $\langle\overline{u_0v_0}\rangle$.   Importantly, this lower order term, associated with ageostrophic vertical motion $w_1$, emerges from  horizontal vorticity dynamics, which is closely linked to eddy momentum convergence; a factor necessary for accelerating the zonal mean zonal wind in the upper atmosphere. 
As will be notable below in our analysis of  the zonally symmetric heat equation \eqref{eq:heat_balance}, this lower order term is associated with eddy-driven mid-latitude jet streams.

In the context of the perturbation expansion presented above, we discern that the anomalous potential temperature $\eta_L$ 
arises from the intricate dynamics of the large-scale atmosphere, which orchestrates the transfer of excess tropical energy to higher latitudes. 
Whereas tropical convection is responsible for negative $\eta_L$, the large-scale mid-latitude dynamics driving poleward eddy heat flux is responsible for the emergence of positive $\eta_L$. The governing equation (\ref{eq:ano_heat}) implies that a steady state results from 
a delicate thermal balance involving the negative feedback from local radiative convective processes, poleward eddy heat flux, 
and adiabatic warming or cooling induced by eddy momentum flux.
%This framework simply expresses the general circulation as being intricately linked to the thermodynamic processes embedded in radiative-convective equilibrium  and the large-scale atmospheric dynamics that generate turbulent eddy heat and momentum fluxes.
 
\section{Extending Held and Hou theory using planetary geostrophic motion\label{sec:HHPG}}

The leading-order dynamics of the Hadley circulation are encapsulated by angular momentum conservation. 
Surface air in the tropics possesses the maximum angular momentum, and in the nearly inviscid limit angular momentum is nearly conserved within the Hadley cell. 
This conservation shapes the zonal mean zonal wind from the tropics to the sub-tropics, and subsequently determines the meridional structure of potential temperature through the thermal wind balance \citep{held1980}.   In this context, we extend the original theory using the formalism of planetary-scale motions.

Whereas the leading-order momentum equation in the mid-latitudes is geostrophic balance, in the tropics  that balance is disrupted because the Coriolis effect becomes negligible. This distinction in the leading-order dynamics underlies the challenges in expressing the dynamical interaction between the tropics and the extra-tropics, which is the origin of using angular momentum conservation as a bridge to dynamically connect the tropics to the mid-latitudes.  

The angular momentum of the large-scale atmosphere is given by $M = \Omega a^2 \cos^2\theta + u^*a\cos\theta$, 
where $\Omega$  and $a$ are the angular velocity and radius of the Earth respectively, $\theta$ is the latitude, and $u^*$ is the zonal velocity. 
In the tropics, the angular momentum of stationary air parcels on the surface is $M=\Omega a^2$, so that the zonal velocity that conserves angular momentum is
\begin{align} \label{eq:um2}
    u^*_M \equiv \frac{\Omega a \sin^2\theta}{\cos\theta} \simeq \Omega a \theta^2, 
\end{align}
where we use the small angle approximation ($\sin\theta \simeq \theta$ and $\cos\theta \simeq 1$). 
Thus, in a zonally symmetric atmosphere, the zonal wind $u^*_M$ can determine the meridional structure of potential temperature through basic balances.

The principal action of the Hadley circulation is to redistribute the low latitude heat surplus towards high latitudes.
Consequently, it is essential to determine the latitudinal distribution of potential temperature, primarily governed by local radiative-convective processes. 
Thus, as discussed above, the focus is on the anomalous potential temperature with respect to the radiative-convective equilibrium potential temperature, {\color{black} the analytical form of which is $\Theta_E = 1-\frac{2}{3}\Delta_H P_2(\sin\theta)$, 
 where $\Delta_H$ is the fractional change in the equator to pole potential temperature, $P_2(x)=\frac{1}{2}(3x^2-1)$ is the second Legendre polynomial.  The thermal Rossby number is
$R=\frac{gH\Delta_H}{\Omega^2 a^2}$, where $H$ the height of the tropical tropopause.
%{\color{red} This can be interpreted as an analytical approximation of the radiative-convective equilibrium temperature.}
%{\color{red}Here, $\Delta_H$ is  
%the fractional change in the equator to pole potential temperature and $P_2(x)=\frac{1}{2}(3x^2-1)$ is the second Legendre polynomial. 
%$R=\frac{gH\Delta_H}{\Omega^2 a^2}$ is the thermal Rossby number with $H$ the vertical length-scale.}
 This leads to the following scaling of the thermodynamic variables; 
 $P^*=P^*_S(z)+\rho^*_s(z)\Omega^2 a^2 P_L$,
$\rho^* = \rho^*_s\left(1+\frac{\Delta_H}{R}\rho_L\right)$ and $\Theta^*=\Theta^*_S(z)(1+\Theta_L)$. }
Finally, the zonal velocity $u^*=\Omega a u_L$ is scaled by $\Omega a$.
This specific scaling is chosen to recover the results as \citet{held1980}.

{\color{black}We use spherical coordinates, and write $X$ and $Y$ as latitude $\theta$ and longitude $\phi$ respectively, so that planetary geostrophic motion under zonally symmetric thermal forcing is given by}
\begin{align}
    &u_L\theta = -\frac{1}{2}\frac{\partial P_L}{\partial\theta}, \label{eq:geostrophy} \\
    &\Theta_L = \frac{\Delta_H}{R}\frac{\partial P_L}{\partial z}, \qquad \text{and} \label{eq:hydro} \\
    &\frac{\partial\eta_L}{\partial \tilde{t}} + u_L\frac{\partial\eta_L}{\partial\phi} =   - \frac{1}{\tau}\eta_L - \frac{\partial}{\partial\theta}\overline{v_0\Theta_0} \label{eq:heat},
\end{align}
where, as previously, the subscripts $L$ and $0$ denote planetary-scale and leading-order synoptic-scale variables respectively.
{\color{black}
Equation (\ref{eq:geostrophy}) is geostrophic balance, wherein we neglect the geometric term $u_L^2\tan\theta$ for simplicity, noting that this does not influence the qualitative results, although
this term can be included for numerical accuracy. Equation (\ref{eq:hydro}) is the thermal wind balance, and the heat equation (\ref{eq:heat})
is written without the contribution of the ageostrophic vertical wind associated with eddy momentum fluxes. Outside of the Hadley cell, 
the influence of synoptic eddies is parameterized by the diffusion of heat and then we impose continuity between the Hadley cell and the mid-latitudes, where we limit our treatment 
to the poleward eddy heat flux.  We note, however, that even when neglecting the eddy momentum flux, the overall results do not change qualitatively. The anomalous warming or cooling induced by the ageostrophic vertical wind is an additional source of heat transfer in the Hadley cell.  
}

%{\color{red}
%Equation (\ref{eq:geostrophy}) is the geostrophic balance in the spherical coordinate.
%we neglect the geometric term $u_L^2\tan\theta$ for simplicity, but note that this does not influence the overall qualitative results,
%and for numerical accuracy this term can be included. Equation (\ref{eq:hydro}) represents thermal wind balance and the last equation (ref{eq:heat})
%is the heat equation without the contribution of ageostrophic vertical wind associated with eddy momentum fluxes. Outside of the Hadley cell, 
%the influence of synoptic eddies is parameterized by the diffusion of heat and then a continuous model between the Hadley cell and the mid-latitudes
%is constructed, for which it is desirable to limit out concern to the poleward eddy heat flux. Even without consideration of the eddy momentum flux, 
%the overall results do not change qualitatively. The anomalous warming or cooling induced by the ageostrophic vertical wind is an additional source
%of heat transfer in the Hadley cell.  
%}
We neglect the meridional, $v_L$, and vertical, $w_L$, velocities due to zonal symmetry, 
and the dynamic mean field depends solely on the zonal wind $u_L$, which satisfies geostrophy (\ref{eq:geostrophy}). 
The key time-evolving equation is the heat equation (\ref{eq:heat}), the zonal average $\langle\cdot\rangle$ of which is
\begin{align}\label{eq:zonal_heat}
    \frac{\partial\eta_L}{\partial \tilde{ t}} = - \frac{1}{\tau}\eta_L - \frac{\partial}{\partial\theta}\langle\overline{v_0\Theta_0}\rangle .
\end{align}
{\color{black} Recall that $v_L = \partial P_L / \partial x$, and hence if there is no longitudinal variation in $P_L$, then $v_L = 0$, 
which implies $\partial (\rho_s w_L) / \partial z = 0$ according to the Sverdrup relation. As a result, $\rho_s w_L$ remains constant throughout the vertical domain, 
making it reasonable to conclude that $w_L = 0$.}
We note that in traditional quasi-geostrophic wave-mean field theory variables are written as the sum of the zonal-averages (the means) and departures from those averages viz. $\theta_0 = \langle\theta_0\rangle+\theta'_0$. 
In the multi-scale formalism, the planetary-scale variables are treated as the means and the synoptic-scale variables the departures from those means, which then have the traditional interpretation in a zonally symmetric atmosphere.

The thermal wind balance obtained by combining equations (\ref{eq:geostrophy}) and (\ref{eq:hydro}) is 
\begin{align}
\label{eq:thwind}
 \frac{\partial u_L}{\partial z}\theta = -\frac{R}{2\Delta_H}\frac{\partial\Theta_L}{\partial\theta},
\end{align}
and integrating both sides in the vertical from 0 to 1 gives 
\begin{align}
u_M \theta = -\frac{R}{2\Delta_H}\frac{\partial\Theta_L}{\partial\theta}.
\label{eq:umtheta}
\end{align}
Using equation (\ref{eq:um2}), where $u^*_M=\Omega a u_M$, we have $u_M \simeq \theta^2$,  so that integration of equation \eqref{eq:umtheta} gives
\begin{align} \label{eq:thetal2}
    \Theta_L(\theta) = \Theta_{\text{eq}} - \frac{\Delta_H}{2R}\theta^4,
\end{align}
where $\Theta_{\text{eq}}$ is the potential temperature at the equator.

\begin{figure}[ht]
\centering
\includegraphics[angle=0,scale=0.20,trim= 0mm 0mm 0mm 0mm, clip]{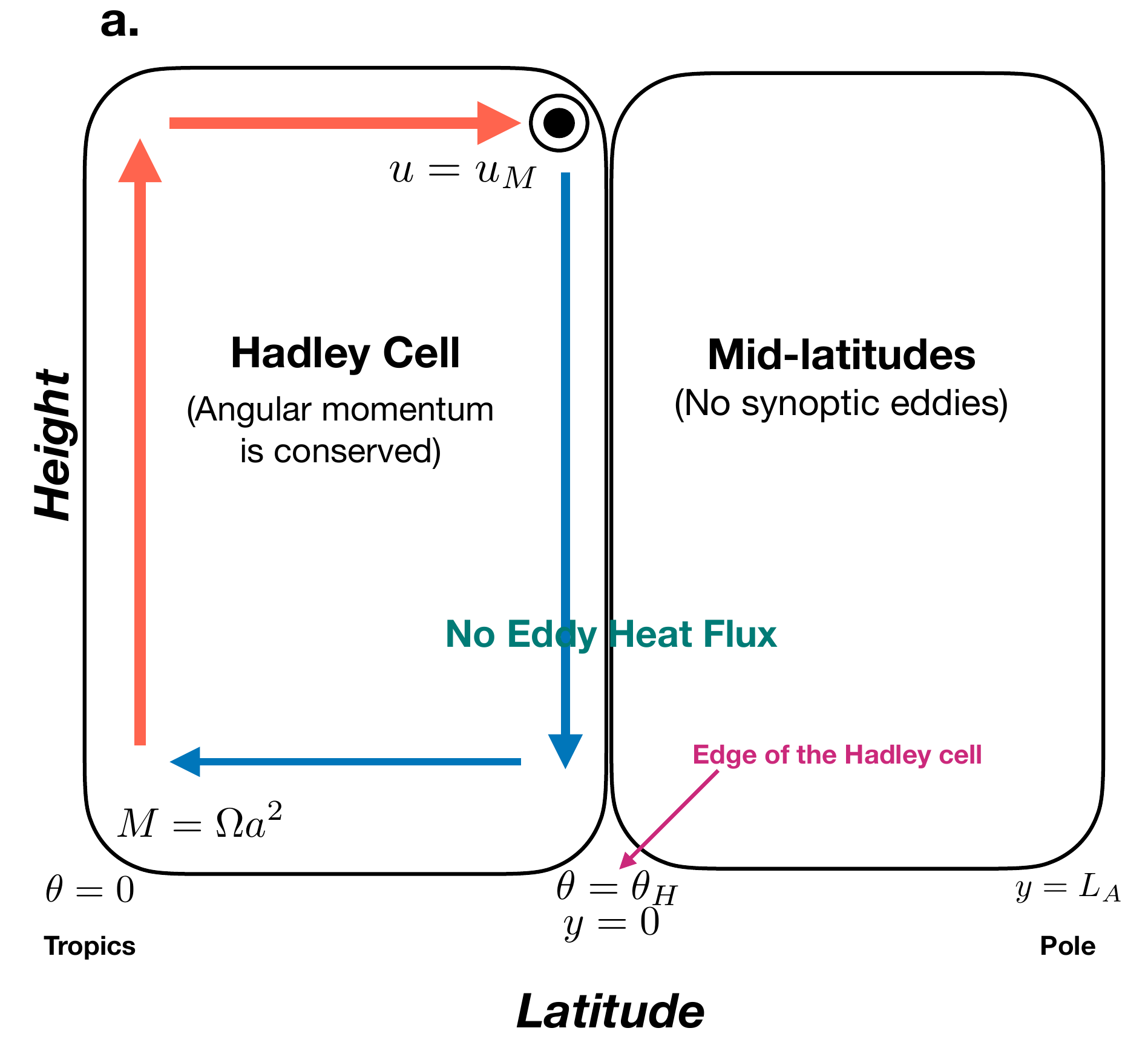}
\includegraphics[angle=0,scale=0.20,trim= 0mm 0mm 0mm 0mm, clip]{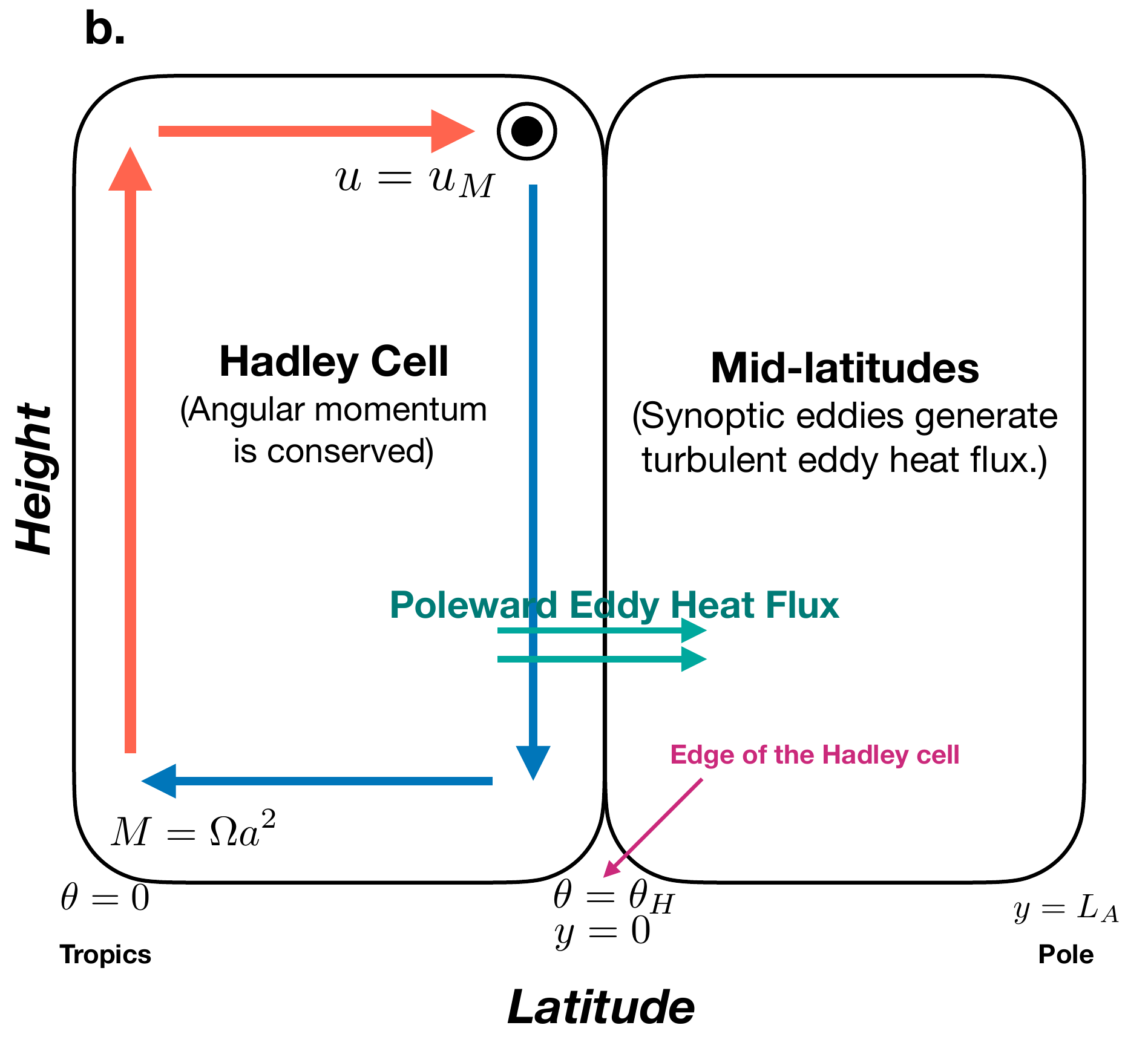}
\caption{(a) A schematic meriodional section of the Hadley circulation theory of \citet{held1980}. Within the Hadley cell the angular momentum of surface air in the tropics is maximal and nearly conserved as it moves upward and poleward towards the edge of the cell at $\theta = \theta_H$. 
It is assumed that the mid-latitude atmosphere is simply defined by a prescribed radiative-convective equilibrium temperature, $\Theta_E$, and the poleward heat flux from tropics to the mid-latitudes is negligible.  (b) A schematic of the simple model described here, which  connects the Hadley cell to the mid-latitudes.  
The principal dynamics within the Hadley cell is same as in \citet{held1980}, but is connected to the mid-latitudes by approximating planetary-scale atmospheric dynamics using 
a diffusive model. }
\label{fig:schematic01}
\end{figure}

The determination of the edge of the Hadley cell ($\theta = \theta_H$) and the equatorial potential temperature ($\Theta_\text{eq}$) 
within the meridional structure of the zonal mean zonal wind and potential temperature is not straightforward. Namely, the challenge associated with localizing the boundaries is due to the ascent of air from the tropics as it moves towards high latitudes, but sinking in the mid-latitudes instead of completing the journey to the polar regions.

Rather than tracking the vertical movement of air,  \citet{held1980} impose two key constraints on potential temperature
to ascertain the cell edge and equatorial potential temperature (Fig.~\ref{fig:schematic01}a). Firstly, continuity of potential temperature demands that $\Theta_L = \Theta_E$ at $\theta = \theta_H$. 
Secondly, for a steady-state Hadley cell, energy flux balance must be maintained. Integrating equation (\ref{eq:zonal_heat}) yields
\begin{align}
    \frac{\partial}{\partial \tilde{t}}\int_{0}^{\theta_f}\eta_L\,d\theta = -\frac{1}{\tau}\int_{0}^{\theta_f}\eta_L\,d\theta - \langle\overline{v_0\Theta_0}\rangle|_{\theta=\theta_f}.
\end{align}

The cell is assumed to be in quasi-isolation, and thus $\langle\overline{v_0\Theta_0}\rangle|_{\theta=\theta_H} \simeq 0$.  
%In the extra-tropics, where baroclinic instability is absent, there is no significant contribution from synoptic eddies in transferring thermal energy poleward.
{\color{black}This case, considered by \cite{held1980}, implies an absence of baroclinic instability in the midlatitudes.}
Thus, the energy flux balance within the cell reduces to
\begin{align}
    \int_{0}^{\theta_H}\Theta_L\,d\theta = \int_{0}^{\theta_H}\Theta_E\,d\theta, 
    \label{eq:thetal}
\end{align}
so that with the prescribed $\Theta_E$ and $\Theta_L$ in equation (\ref{eq:thetal}), the edge of the Hadley cell is given as
\begin{align}
    \theta_H = \left(\frac{5}{3}R\right)^{1/2}=\left(\frac{5}{3}\right)^{1/2}\left(\frac{gH\Delta_H}{\Omega^2a^2}\right)^{1/2}.
\end{align}

The continuity of potential temperature and the steady-state energy flux balance are the two essential assumptions that form the basis of the \citet{held1980} theory.
The resulting values of  $\theta_H$ and $\Theta_0$ define the meridional profiles of potential temperature $\Theta_L$, and thus determine the poleward heat flux within the cell $\langle\overline{v_0\Theta_0}\rangle$.

Assuming self-similarity of the synoptic scale potential temperature $\Theta_0$ and zonal wind $u_0$, one can deduce the meridional structure of the momentum flux $\langle\overline{u_0v_0}\rangle$ 
and the surface wind, providing the easterlies in the tropics.  We note again that the small-angle approximation is used for simplicity.  A comprehensive treatment involves use of full spherical coordinates, but our aim here is to provide general qualitative insight.

\subsection{The interaction of the Hadley circulation with the mid-latitudes\label{sec:inter-model}}

The \citet{held1980} theory does not treat the influence of mid-latitude synoptic eddies. In the vicinity of the Hadley cell edge, the subtropical jet 
undergoes baroclinic instability, initiating the life cycle of synoptic eddies. Baroclinic instability results in the amplification of synoptic waves, 
causing lower-level poleward heat flux. Saturated upper-level synoptic eddies break on the equatorward flank of the jet stream, 
generating eddy momentum flux convergence and intensifying the zonal mean zonal wind.  Synoptic eddy breakage creates downward motion associated with the eddy momentum flux and thus drives adiabatic warming.  

%The intricate dynamics of synoptic eddies underlie the transfer of the heat surplus from the Hadley cell to the mid-latitudes. 
The two constraints of the \citet{held1980} theory used to determine the potential temperature at the equator and the edge of the Hadley cell discussed above, originate in the heat equation.  
The main consequence of neglecting the energy transfer by synoptic eddies near the Hadley cell edge is the equal area constraint of equation (\ref{eq:thetal}), 
stating that the latitudinal integral of potential temperature equals that of the radiative convective equilibrium.  However, treating mid-latitude synoptic eddy activity is synonymous with accounting for heat transfer from the Hadley cell to the mid-latitudes, which drives a non-zero mid-latitude anomalous potential temperature, and this must be determined in a manner consistent with  
continuity of potential temperature at the Hadley cell edge.

 \subsubsection{The Hadley cell}
 
 Angular momentum conservation determines the dynamic and thermodynamic structure of the Hadley cell as in \cite{held1980}.  
 Consequently,  the upper-level zonal mean zonal wind $u^*_M$ is given by equation (\ref{eq:um2}).  Hence,  as shown above, using the thermal wind balance of equation \eqref{eq:thwind} the potential temperature is given by equation (\ref{eq:thetal2}).  
Therefore, the anomalous potential temperature $\eta_L(\theta)$ is 
\begin{align}
\label{eq:PTtropics}
\eta_L (\theta) = \Theta_\text{eq}-\left(1+\frac{1}{3}\Delta_H\right)+\Delta_H\theta^2 -\frac{\Delta_H}{2R}\theta^4
\end{align}
The constant of integration is the potential temperature at the equator, $\Theta_\text{eq}$,  which remains undetermined at the level of the thermal wind balance; it is determined by treating the interaction between the Hadley cell and the mid-latitudes through the heat flux balance, as discussed next.

\subsubsection{The Mid-latitudes}

The mid-latitudes range from the northern edge of the Hadley cell,  $\theta=\theta_H$, to the North Pole, $\theta=\theta_f$. 
Thus, extension of the \citet{held1980} theory to incorporate the influence of synoptic eddies near the high-latitude edge of the Hadley cell requires the determination of 
the meridional distribution of the anomalous potential temperature throughout the mid-latitudes as shown in Fig.~\ref{fig:schematic01}(b).
The primary objective is to account for the impact of the Hadley cell to mid-latitude eddy energy flux, which generates positive mid-latitude anomalous potential temperature.
To obey the constraint of potential temperature continuity at the Hadley cell edge, a simple mid-latitude thermodynamic model is needed to redistribute the energy received from the Hadley cell.

We emphasize that our aim is not to  construct an analytical model  that captures realistic mid-latitude circulation.   Rather, here we provide a simplified treatment of the synoptic eddy driven energy flux between the Hadley cell and the mid-latitudes, with a particular emphasis on its impact on the intensity and size of the Hadley cell.  For simplicity, as discussed above, we make the $\beta$-plane approximation and use a Cartesian meridional coordinate system, with $y=0$ ($y=L_A$) corresponding to $\theta=\theta_H$ ($\theta=\theta_f$), as shown in Fig.~\ref{fig:schematic01}(b).  We use $y$ as the independent variable in the mid-latitudes and $\theta$ in the tropics and match these appropriately as described below. 
Thus, we rewrite the zonally symmetric heat equation \eqref{eq:zonal_heat} as 
\begin{align} \label{eq:heat_balance}
 \frac{\partial\eta_L}{\partial \tilde{t}}& = -\frac{1}{\tau}\eta_L-\frac{\partial}{\partial y}\langle\overline{v_0\Theta_0}\rangle \nonumber\\
& \equiv  -\frac{1}{\tau}\eta_L +  D\frac{\partial^2\eta_L}{\partial y^2}, 
\end{align}
which as noted above neglects ageostrophic vertical motion and thus provides a very simple model for redistributing heat flux originating from the Hadley cell.  
Namely, the poleward heat flux is assumed to be Fickian and hence $\langle\overline{v_0\Theta_0}\rangle = -D\frac{\partial\eta_L}{\partial y}$
with an eddy diffusivity $D$.  The Fickian approximation for the poleward eddy heat flux is discussed extensively  by \citet{held2000} 
and has been validated against observational reanalysis data \citep{barry2002}.  Moreover, the Fickian approximation can be modified to incorporate the eddy memory effect, known as the minimal $\tau$ approximation, thereby capturing the quasi-oscillatory behavior of the poleward heat flux in large-scale mid-latitude dynamics \citep{moon2021}.
Clearly, this approximation is valid when planetary variables evolve much more slowly than the synoptic eddy turnover time-scale, and is thus appropriate in treating steady state poleward eddy heat flux, for which equation \eqref{eq:heat_balance} becomes
\begin{align}
\label{eq:ss}
 D\frac{\partial^2\eta_L}{\partial y^2}-\frac{1}{\tau}\eta_L=0, 
\end{align}
and is governed by the following boundary conditions, as shown in Fig.~\ref{fig:schematic01} (b).  The poleward edge of the Hadley cell is at $y=0$ ($\theta=\theta_H$), and the North Pole is at $y=L_A$ ($\theta=\pi/2$).
The boundary condition at $y=0$ is that $\langle\overline{v_0\Theta_0}\rangle$ is continuous, and hence
\begin{align}
\lim_{\theta\to\theta_H-}\frac{\partial\eta_L}{\partial\theta}=\lim_{y \to 0+ (\theta\to\theta_H+)}\frac{\partial\eta_L}{\partial y}.
\end{align}
The meridional heat flux vanishes at the North Pole, so that
\begin{align}
-D\frac{\partial\eta_L}{\partial y}|_{y=L_A} = 0.
\end{align}

\subsubsection{Interaction between the tropics \& mid-latitudes}

The anomalous potential temperature in the tropics is given by equation \eqref{eq:PTtropics}, and in the mid-latitudes the solution to equation \eqref{eq:ss} is $\eta_L(y) = K_1e^{y/\sqrt{D\tau}}+K_2e^{-y/\sqrt{D\tau}}$,
with two undetermined constants $K_1$ and $K_2$. The boundary conditions at $y=0$ and $y=L_A$ give
\begin{align}
 -K_1+K_2 &= 2\Delta_H\sqrt{D\tau}\left(\frac{\theta^3_H}{R}-\theta_H\right) \qquad \text{and} \\
 K_2 &=K_1\text{exp}\left(2\frac{L_A}{\sqrt{D\tau}}\right),
\end{align} 
from which we find
\begin{align}
&K_1=\frac{2\Delta_H\sqrt{D\tau}}{e^{2\frac{L_A}{\sqrt{D\tau}}}-1}\left(\frac{\theta^3_H}{R}-\theta_H\right) \qquad \text{and} \\
&K_2=2\Delta_H\sqrt{D\tau}\frac{e^{2\frac{L_A}{\sqrt{D\tau}}}}{e^{2\frac{L_A}{\sqrt{D\tau}}}-1}\left(\frac{\theta^3_H}{R}-\theta_H\right).
\end{align}

To uniquely determine the meridional structure of the anomalous potential temperatures $\eta_L$, we require two constraints to determine the 
two unknowns $\Theta_\text{eq}$ and $\theta_H$. The first is the continuity of potential temperature, 
\begin{align}\label{eq:const1}
\lim_{\theta\to\theta_H-}\eta_L=\lim_{\theta\to\theta_H+}\eta_L, 
\end{align}
and the second is energy flux balance in the Hadley cell. Integrating \eqref{eq:ss} with respect to  $\theta$ gives
\begin{align}\label{eq:const2}
D\frac{\partial\eta_L}{\partial\theta}|_{\theta=\theta_H}-\frac{1}{\tau}\int_{0}^{\theta_H}\eta_L d\theta = 0, 
\end{align}
which when combined with \eqref{eq:const1} gives 
\begin{align}\label{eq:thetah}
&2\Delta_H\sqrt{D\tau}\text{coth}\left(\frac{L_A}{\sqrt{D\tau}}\right)\left(\frac{\theta^3_H}{R}-\theta_H\right) \nonumber \\
&=-\frac{2\Delta_H}{5R}\theta^4_H+\frac{2}{3}\Delta_H\theta^2_H-2D\tau\Delta_H\left(\frac{\theta^2_H}{R}-1\right),
\end{align}
and
\begin{align}\label{eq:theta0}
\Theta_\text{eq} &=\frac{\Delta_H}{10R}\theta^4_H-\left(\frac{2D\tau\Delta_H}{R} 
+\frac{1}{3}\Delta_H\right)\theta^2_H \nonumber \\
&+\left(1+\frac{1}{3}\Delta_H\right)+2D\tau\Delta_H.
\end{align}
Therefore, equations (\ref{eq:thetah}) and (\ref{eq:theta0}) determine the two unknown coefficients $\theta_H$ and $\Theta_\text{eq}$.

Integrating the stationary form of the first version of heat flux balance in \eqref{eq:heat_balance} 
yields the meridional structure of the poleward heat flux $\overline{v_0\Theta_0}$ within the Hadley cell as 
\begin{align} 
\label{eq:HCflux}
\overline{v_0\Theta_0} &=-\frac{1}{\tau}\int_{0}^{\theta}\eta_L d\theta^\prime \nonumber \\
&=-\frac{1}{\tau}\left[\Theta_\text{eq}\theta-\left(1+\frac{1}{3}\Delta_H\right)\theta+\frac{1}{3}\Delta_H\theta^3-\frac{\Delta_H}{10R}\theta^5\right].
\end{align}
On the other hand, in the mid-latitudes ($\theta > \theta_H$) we have
\begin{align}
\label{eq:MLflux}
 \overline{v_0\Theta_0} = -D\frac{\partial\eta_L}{\partial y}=-\sqrt{\frac{D}{\tau}}\left(K_1e^{\frac{y}{\sqrt{D\tau}}}-K_2e^{-\frac{y}{\sqrt{D\tau}}}\right).
\end{align}

Akin to \citet{held1980}, to obtain an estimate of the momentum fluxes, we assume that the profiles of $u_0$ and $\Theta_0$ are self-similar, which leads to
\begin{align}
\overline{u_0v_0} \simeq \frac{u_L(z=1)-u_L(z=0)}{\Theta_L(z=1)-\Theta_L(z=0)}\overline{v_0\Theta_0} \simeq \frac{\theta^2}{\Delta_v}\overline{v_0\Theta_0},
\end{align}
where $\Delta_v \equiv \Theta_L(z=1)-\Theta_L(z=0)$ represents the vertical stability of atmosphere. 
From the zonal momentum equation,  the surface stress in the tropics can be deduced as
\begin{align} \label{eq:cutr}
\frac{\partial}{\partial\theta}\overline{u_0v_0} \simeq -Cu_L(z=0), 
\end{align}
and similarly in the mid-latitudes,
\begin{align}  \label{eq:cuml}
\frac{\partial}{\partial y}\overline{u_0v_0} \simeq -Cu_L(z=0),
\end{align}
from which the trade wind and the mid-latitude surface wind are calculated respectively, and $C$ is a constant drag coefficient. 

\section{Poleward heat flux and the Hadley cell dynamics}

Equation \eqref{eq:thetah} gives the poleward edge of the Hadley cell, $\theta_H$, as a function of $D\tau$ and the thermal Rossby number $R$, which is proportional to $H$, the height of the tropical tropopause as noted previously. This height is controlled by tropical radiative-convective equilibrium in the upper troposphere, stratospheric circulation, and the interaction with mid-latitude
eddies \citep[e.g.,][]{fueglistaler2009}, and thus may be linked, directly or indirectly, to recent global warming. The eddy diffusivity $D$ is related to baroclinic
instability of the subtropical jet stream, which generates poleward heat transport through synoptic eddies \citep{north1975, taylor1980}, 
and $\tau^{-1}$ is proportional to the overall climate sensitivity \citep{moon2021}. Therefore, $\tau$
it is closely related to the magnitude of positive feedbacks in climate system \citep{moon2017}.

{\color{black} Under global warming, the increase in surface temperature varies geographically. In particular, the polar regions experience a greater increase in surface temperature, 
which can influence the mid-latitude baroclinicity, creating a possible causal link between global warming and the eddy diffusivity, $D$.
However, this linkage may involve more than the mean baroclinicity. 
Several studies suggest that the dominant eddy length scale increases in response to global warming \citep{geen2023, xue2017, chemke2020L}. 
In particular, \citet{chemke2020L} found that while the synoptic-scale eddy kinetic energy decreases under global warming, 
it increases on the planetary scale. They emphasized that this shift is a result of the change in the wave-mean interactions. 
Furthermore, it has been proposed that baroclinic instability on the planetary scale requires an optimal level of baroclinicity, 
as opposed to the stronger baroclinicity needed for classical synoptic-scale baroclinic instability \citep{moon2022}. 
Clearly, the interesting question of how $D$ evolves across a range of climate states requires further investigation.
}

The interannual variability of the height of the tropical tropopause is $50$ m \citep{randel2000}, and has not shown an increase sufficient to change the size of the Hadley cell  \citep{seidel2006}.  Therefore, the influence of global warming on the size of the Hadley cell may be reflected through the eddy diffusivity $D$ and the response time scale $\tau$. It is known that mid-latitude baroclinic eddies can control the size and intensity of the Hadley cell \citep{walker2006, levine2015}, and these eddies are generated by jet stream instabilities \citep{pierrehumbert1995}.  Thus, there is substantial baroclinic eddy activity of the subtropical jet near the edge of the Hadley cell.

Therefore, the eddy diffusivity (a) represents the role of synoptic eddies on the poleward heat transfer; particularly near the poleward edge of the Hadley cell, and (b) has a magnitude 
closely related to the planetary baroclinicity, which is proportional to the meridional potential temperature gradient \citep{barry2002}.
Although the large-scale atmospheric dynamics of the subtropical jet is too complicated to be fully described
by a constant $D$ with meridional temperature gradient,  the overall impact of synoptic eddies on the Hadley cell may be captured 
by an eddy diffusivity \citep{north1981, barry2002}.

\begin{figure}
\centering
\includegraphics[angle=0,scale=0.12,trim= 0mm 0mm 0mm 0mm, clip]{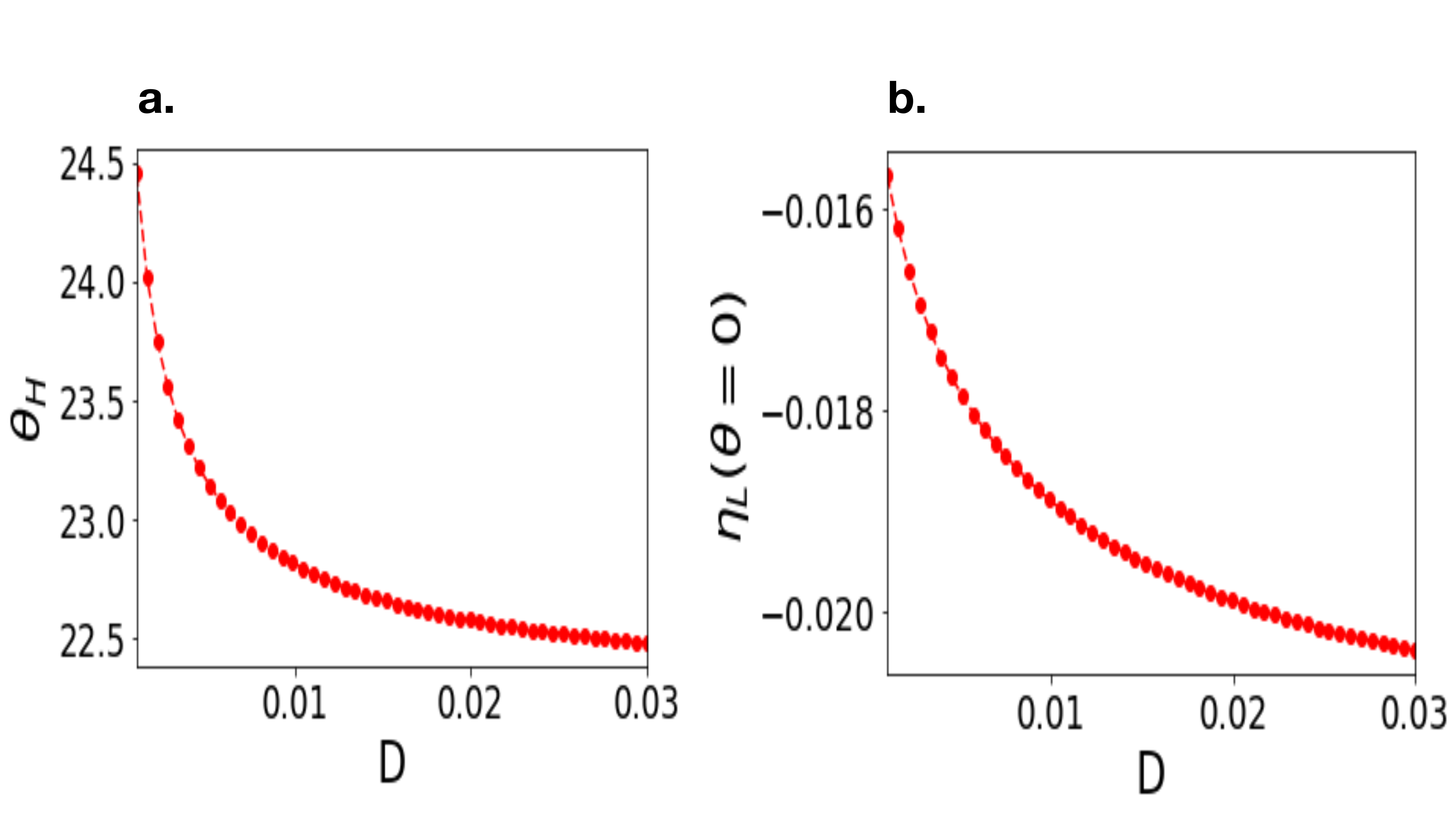}
\caption{The change in (a) the position of the Hadley cell $\theta_H$, and (b) the 
anomalous potential temperature $\eta_L(\theta=0)$ at the equator, as a function of the eddy diffusivity $D$. 
 }
\label{fig:fig02}
\end{figure}

Reanalysis data and model simulations show that the mid-latitude eddy heat flux decreases under the influence of the Arctic Amplification \citep{chemke2020}, implying that $D$ changes with global warming, and hence in the context of our model can lead to a change in the Hadley circulation.  
{\color{black} According to \citet{barry2002}, the eddy heat flux can be approximated as $-D \partial \Theta_L / \partial y$, 
where $D$ is a function of the mean baroclinicity. Consequently, changes in baroclinicity due to Arctic Amplification are reflected in a reduction of the eddy diffusivity $D$.}
We fix $\tau$ and $R$ and solve \eqref{eq:thetah} for a range of eddy diffusivities.  For planetary geostrophic motion, \citet{moon2021} estimated the dimensionless synoptic eddy diffusivity as $D\simeq 1/5\pi^2 \sim 0.03$, to explain the quasi-oscillatory behavior of the baroclinic annular mode. 

Here we vary $D$ from $0.001$ to $0.03$ to observe the effect of decreasing the poleward 
heat flux in the mid-latitudes under global warming.  In Fig. \ref{fig:fig02} (a) we show that the poleward edge of the Hadley cell, $\theta_H$, moves poleward as $D$ decreases, changing rapidly for small $D$ and asymptoting for larger values.   It is expected that the Hadley cell moves further poleward during summer when the baroclinicity is weak \citep{hu2018}.   In Fig. \ref{fig:fig02} (b)  we see that the negative anomalous potential temperature at the equator, $\eta_L(\theta=0)$, becomes more negative as $D$ increases. 

\begin{figure}
\centering
\includegraphics[angle=0,scale=0.19,trim= 0mm 0mm 0mm 0mm, clip]{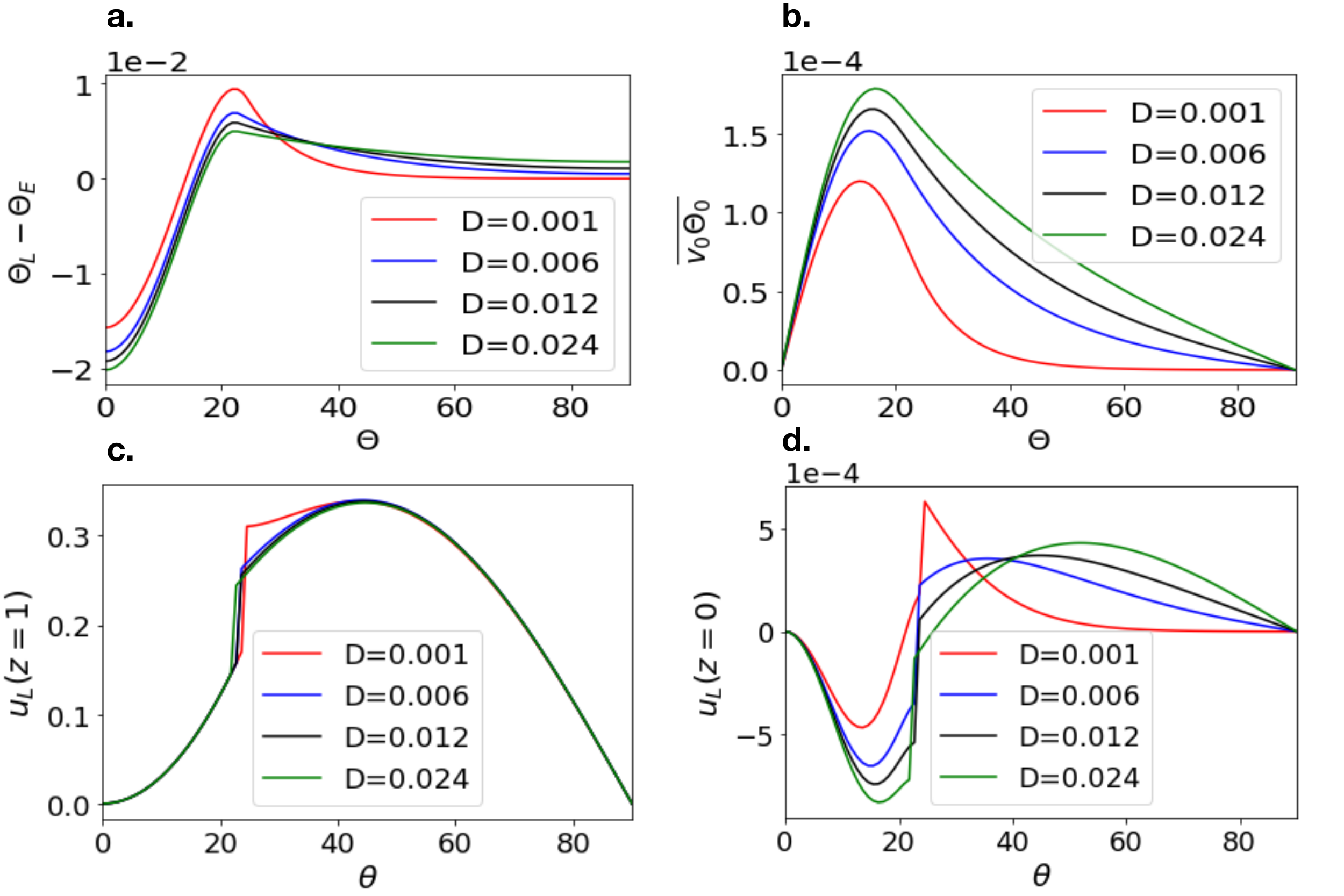}
\caption{For the  range of eddy diffusivities $D$, shown in the insets, we plot (a) the anomalous potential temperature $\eta_L=\Theta_L - \Theta_E$;  (b) the poleward heat flux, from equations \eqref{eq:HCflux} and \eqref{eq:MLflux}; (c) upper level, $u_L(z=1)$,  and (d) the lower level, $u_L(z=0)$,  zonal mean zonal wind, from the tropics to the pole. 
In the Hadley cell ($0 \le \theta \le \theta_H$), $u_L(z=1)$= $u_M$, whereas outside of the Hadley cell  it is calculated from the thermal wind balance.
The lower level wind, $u_L(z=0)$, is determined from equations \eqref{eq:cutr} and \eqref{eq:cuml}. 
}
\label{fig:fig03}
\end{figure}

Fig. \ref{fig:fig03}(a) shows the meridional structure of the anomalous temperature from the tropics poleward; negative in the tropics, increasing to a positive maximum at the edge of the Hadley cell, and then slowly decreasing towards the polar regions. When $D$ is smaller (larger), $\eta_L$ at the edge of the cell becomes larger (smaller) and decreases rapidly (slowly) with increasing latitude. Fig. \ref{fig:fig03}(b) shows that the poleward heat flux, $\overline{v_0\Theta_0}$, is maximal in the middle of the Hadley cell and decreases in the mid-latitudes. As expected, for smaller $D$ the overall magnitude of the heat flux is smaller.

Following the thermal wind balance, the meridional structure of the potential temperature determines the zonal mean zonal wind $u_L$.  Fig. \ref{fig:fig03}(c) shows the upper level zonal mean zonal wind $u_L(z=1)$, which decreases at the edge of the Hadley cell as $D$ increases.  The surface zonal mean zonal wind, $u_L(z=0)$, shown in Fig. \ref{fig:fig03}(d), reflects the intensity of the Hadley cell and the magnitude of $\eta_L$ at the equator, measured by the magnitude of the tropical trade winds. As $D$ increases, the surface easterlies in 
 tropics strengthen. This implies that the increase in the poleward heat flux by synoptic eddies in mid-latitudes
 can control the intensity of the Hadley cell. As the extent of the Hadley cell decreases (increases) 
 due to the increase (decrease) of the poleward heat
 flux, the intensity of the cell increases (decreases). 
 
 \section{\label{sec:discussion} Discussion}
 
We have extended the Hadley circulation model of \citet{held1980} by including a simple extratropical thermodynamic treatment that reveals the linkage between the tropics and the mid-latitudes. 
In particular, we find that the poleward heat flux driven by baroclinic instability near the poleward edge of the Hadley cell can change 
the dynamical features of the Hadley circulation, including the meridional extent of the cell. 

The role of the poleward heat flux at the edge of the cell can be considered as follows.  
In the limit $D \ll 1$, let $\delta \equiv \sqrt{D\tau}$ and consider the perturbative solution
$\theta_H \simeq \theta^0_H+\delta\theta^1_H$ of \eqref{eq:thetah}, which leads to
$\theta^0_H =\left(\frac{5}{3}R\right)^{1/2}$ and $\theta^1_H = -1$. This is equivalent to
$\theta_H \simeq \left(\frac{5}{3}R\right)^{1/2}-\sqrt{D\tau}$. Therefore, for $R$ fixed,  
the deviation is $\Delta\theta_H \simeq -\sqrt{D\tau}$.

 The poleward heat flux at the edge of the Hadley cell $\overline{v_0\Theta_0}|_{\theta=\theta_H}$ is proportional 
 to the degree of baroclinic instability of the subtropical jet.   This can be measured by the growth rate of baroclinic waves, 
 which according to the Eady problem is $\sigma_{max} = 0.31\frac{f}{NH_e}\Delta \overline{U}$, where 
 $N$ is the static stability, $H_e$ the tropopause height at the edge of the cell, and $\Delta \overline{U}$ is the shear of the zonal mean
 zonal wind. The poleward heat flux is generated during the growth stage of synoptic waves by baroclinic instability. 
Hence, $\overline{v_0\Theta_0}|_{\theta=\theta_H} \propto \sigma_{max}$, so that $D \propto  \overline{v_0\Theta_0}|_{\theta=\theta_H}  \propto 1/H_e$.
Therefore, the width of the Hadley cell decreases as $\sqrt{D\tau}$ increases, and the increase of $H_e$ leads to the decrease in $D$, 
implying that the width of the Hadley cell increases. 
 
\citet{lu2007} considered a scaling of $\theta_H$ suggested by \citet{held2000}, 
wherein one defines the poleward edge of the Hadley cell as the latitude at which the vertical shear of the zonal wind 
is baroclinically unstable, giving $\theta_H \propto \left(\frac{NH_e}{\Omega a}\right)^{1/2}$.
It is known that $H_e$ is highly correlated with the expansion of the Hadley cell \citep{lu2007}, and that baroclinic eddy fluxes 
are central in determining the extratropical tropopause height and thermal stratification \citep{schneider2004}. 
Therefore, turbulent synoptic motions initiated by baroclinic
instability due to the vertical shear of the zonal wind near the edge of the cell are crucial in determining the scaling of $\theta_H$.

Here, we directly estimate the change of the Hadley cell size due to the degree of baroclinicity of the subtropical jet. 
As the tropopause height $H_e$ increases and the vertical shear of the jet decreases, 
the size of the Hadley cell increases. This is consistent with the scaling used by \citet{lu2007}, however
our scaling comes from an exact extension of the original theory that includes interaction with the mid-latitudes 
rather than a phenomenological conjecture. 

Global climate model simulations show that during El Ni\~{n}o events the Hadley circulation strengthens and contracts toward tropics, whereas under global warming 
it weakens and expands poleward \citep{lu2008}. During El Ni\~{n}o, the heat is concentrated in the tropics and 
baroclinicity near the edge of the Hadley cell increases. The theory described here suggests the same intensification and contraction. However, under 
global warming the temperature in the mid-latitudes increases more than in the tropics, which leads to a decrease in baroclinicity near the edge
of the Hadley cell. Idealized dynamic core simulations also show that the enhanced surface heat flux in the tropics 
leads to the intensification of the Hadley cell \citep{son2005}. Reanalysis data and climate model simulations consistently 
show a weakening and widening of the Hadley circulation due to a decrease in mid-latitude temperature gradients \citep{adam2014}.
These findings are consistent with the theory presented here. 
{\color{black}It is assumed that the degree of baroclinic instability is primarily governed by the vertical shear of the lower level jet stream. 
Previous research suggests that, under global warming, the vertical shear of the upper level jet stream increases \citep{lorenz2007}, 
which may contrast with the expected decrease in eddy diffusivity, $D$, under global warming. 
However, classical baroclinic instability theory indicates that the mid-latitude eddy heat flux is maximized 
near the surface \citep{pierrehumbert1995}, implying that the overall eddy heat flux is predominantly influenced by the lower level baroclinicity.}

Our rationale for neglecting the influence of the eddy momentum flux in the heat equation is as follows.  The steady state planetary-scale equations are
\begin{align}
 &-\frac{\partial}{\partial\theta}\langle\overline{v_0\Theta_0}\rangle-\frac{S}{\epsilon^{1/2}}\langle\overline{w_1}\rangle-\frac{1}{\tau}\eta_L=0. \qquad \text{and}  \label{eq:steady1} \\
 &\left(\frac{\partial}{\partial z}-\frac{1}{H}\right)\langle\overline{w_1}\rangle = -\epsilon\frac{\partial^2}{\partial\theta^2}\langle\overline{u_0v_0}\rangle, \label{eq:steady2}
\end{align}
where the eddy momentum flux is represented by the averaged vertical velocity $\langle\overline{w_1}\rangle$, and Eq. \eqref{eq:steady2} is 
derived from the temporal and zonal mean of the quasi-geostrophic potential vorticity equation.
The small planetary-scale Rossby number $\epsilon$ 
arises from the spatial average of the synoptic derivatives, and $-\frac{1}{H} =\frac{1}{\rho_S}\frac{\partial\rho_S}{\partial z}$, with
$H$ the density scale height, assumed to be constant \citep{moon2020}.
Consequently, $\frac{S}{\epsilon^{1/2}}\langle\overline{w_1}\rangle \sim O(\epsilon^{1/2})$, and the contribution of the vertical velocity is asymptotically smaller 
than that of the eddy heat flux, which is the key point we develop here, by revisiting the energy flux balance inside the Hadley cell.  
Integrating Eqs. \eqref{eq:steady1} and \eqref{eq:steady2} and eliminating the averaged vertical velocity we obtain
\begin{align}
& \frac{1}{\tau}\int_{0}^{1}\int_{0}^{\theta_H}\eta_Ld\theta dz = -\int_{0}^{1}\langle\overline{v_0\Theta_0}\rangle|_{\theta_H}dz \nonumber \\
& -\epsilon^{1/2}SH\int_{0}^{1}\frac{\partial}{\partial\theta}\langle\overline{u_0v_0}\rangle |_{\theta_H}dz,
 \label{eq:intbalance}
\end{align} 
using $w_1=0$ at $z=0$ and $z=1$.

The excess energy in the Hadley cell is transported to the high latitudes, where we assume that the eddy heat and momentum fluxes become negligible, and hence we have
\begin{align}
\int_{0}^{1}\int_{0}^{\pi/2}\eta_L d\theta dz = 0,
\end{align} 
so that
\begin{align}
\int_{0}^{1}\int_{0}^{\theta_H}\eta_Ld\theta dz = -\int_{0}^{1}\int_{\theta_H}^{\pi/2}\eta_Ld\theta dz.
\end{align}
Because the surplus energy in the tropics is transported to the high latitudes, the high latitude integrals of $\eta_L$ 
should be positive, and hence
\begin{align}\label{eq:constraint}
\int_{0}^{1}\int_{0}^{\theta_H}\eta_Ld\theta dz < 0.
\end{align}
Therefore, because (a) the eddy heat flux $\langle\overline{v_0\Theta_0}\rangle$ in mid-latitudes generated by baroclinic instability is positive, implying that
synoptic waves are generated to transfer the surplus of energy to high latitudes, and (b) the eddy momentum flux changes
from positive to negative when transecting the jet stream, so that $\frac{\partial}{\partial\theta}\langle\overline{u_0v_0}\rangle < 0$, then 
\begin{align}
-\int_{0}^{1}\langle\overline{v_0\Theta_0}\rangle|_{\theta_H}dz &< 0 \qquad \text{and} \nonumber \\
 -\epsilon^{1/2}SH\int_{0}^{1}\frac{\partial}{\partial\theta}\langle\overline{u_0v_0}\rangle |_{\theta_H}dz &> 0.   
\end{align}
Taken together with the constraint \eqref{eq:constraint}, we have 
\begin{align}
\left| \int_{0}^{1}\langle\overline{v_0\Theta_0}\rangle|_{\theta_H}dz \right| > 
\left| \epsilon^{1/2}SH\int_{0}^{1}\frac{\partial}{\partial\theta}\langle\overline{u_0v_0}\rangle |_{\theta_H}dz \right|, 
\label{eq:ineq}
\end{align}
so that the influence of the eddy heat flux on the expansion or contraction of the Hadley cell is modulated by the eddy momentum flux. This is consistent
with the life cycle of mid-latitude baroclinic waves. The poleward heat flux generated by baroclinic waves weakens the jet stream, 
but the eddy momentum flux induced by wave breaking near the critical latitudes strengthens the jet stream. 

{\color{black} Previous research highlights the role of synoptic eddies through various eddy metrics, such as eddy momentum flux divergence, supercriticality, 
and dry static stability \citep{walker2006, korty2008, levine2015}. These metrics are closely linked to the baroclinic wave life cycle, which begins with baroclinic instability 
generating eddy heat flux near the surface. Within the framework of the planetary-scale heat equation, these eddy properties are associated 
with heat transfer processes in the Hadley cell.
On the one hand, eddy momentum flux (or its divergence) and dry static stability determine adiabatic warming, 
contributing to the expansion of the Hadley cell. On the other hand, supercriticality and baroclinicity regulate eddy heat flux, 
which can lead to a contraction of the Hadley cell.
By considering the baroclinic wave life cycle and its role in the energy flux balance of the Hadley cell, 
our framework aligns with these previous studies that emphasize the influence of synoptic eddies 
on the size and intensity of the Hadley cell.}

The essential point is that the magnitude of the poleward heat flux at the edge of the Hadley cell controls its size and intensity.  Clearly, the inequality \eqref{eq:ineq} could be divined from the small $\epsilon$ limit of \eqref{eq:intbalance}, but the role of the eddy heat flux is the same.   However, a precise calculation of the size of the Hadley cell requires treating both the eddy momentum flux and the eddy heat flux.

\begin{figure}
\centering
\includegraphics[angle=0,scale=0.33,trim= 10mm 35mm 10mm 10mm, clip]{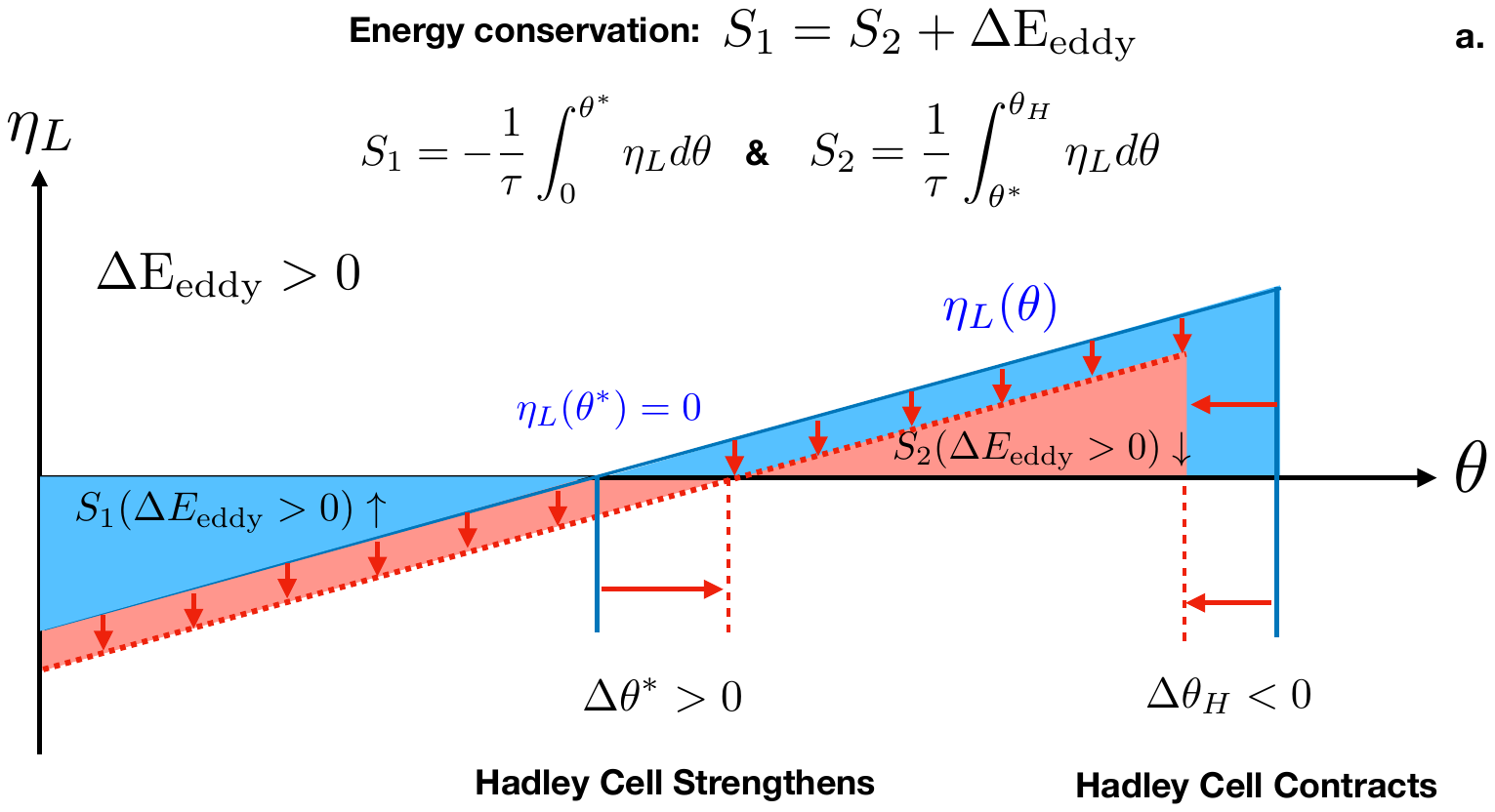}\\
\includegraphics[angle=0,scale=0.33,trim= 10mm 35mm 10mm 50mm, clip]{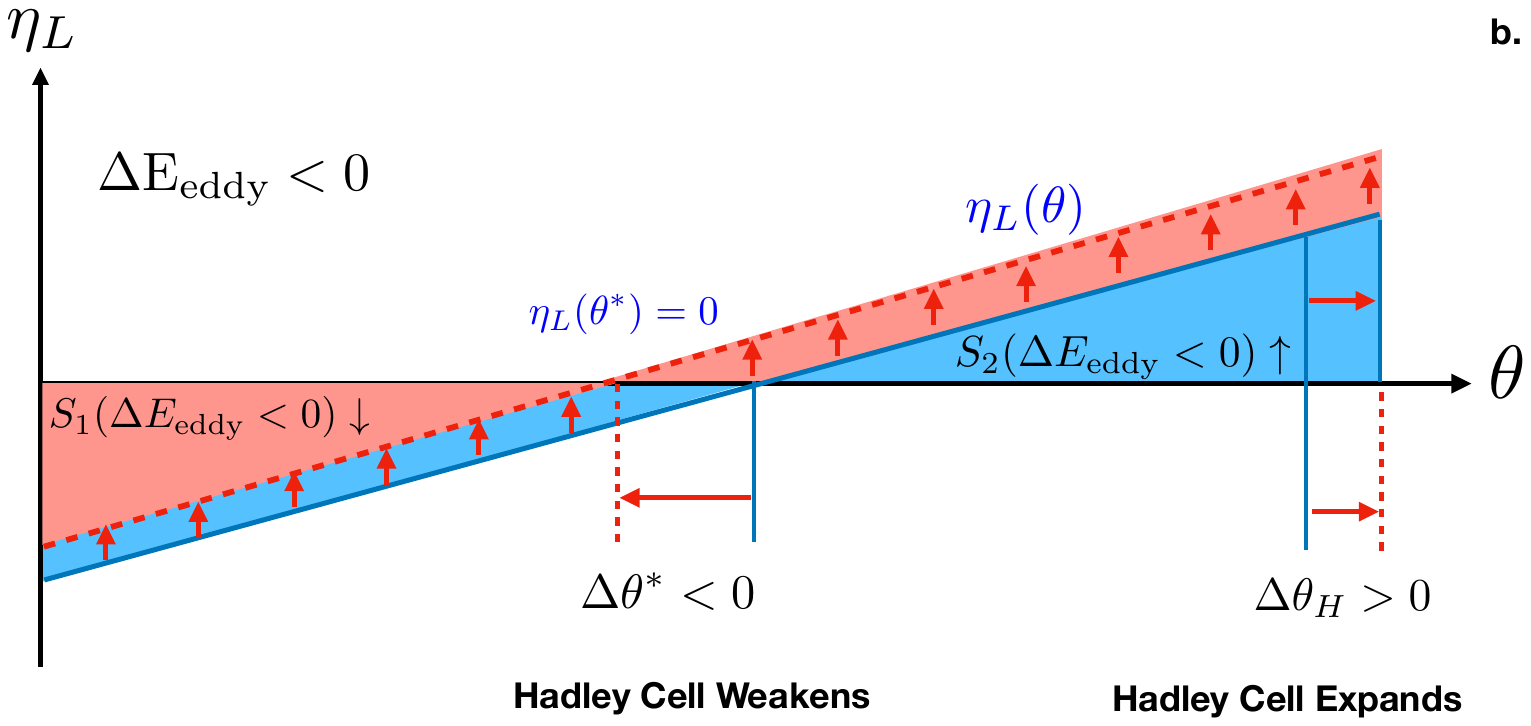}
\caption{Schematic showing the impact of changing the poleward eddy heat flux, $\Delta\text{E}_\text{eddy}$, at the edge of the Hadley cell, $\theta_H$. {\color{black}
Hemispheric thermal energy conservation requires $S_1=S_2+\Delta\text{E}_\text{eddy}$, and the figure shows the changes in $S_1=S_1(\Delta\text{E}_\text{eddy})$ and $S_2=S_2(\Delta\text{E}_\text{eddy})$, indicated by the red arrows, as $\Delta\text{E}_\text{eddy}$ (a) increases or (b) decreases (see text for more discussion).  
An increase (decrease) in the poleward heat flux $\Delta\text{E}_\text{eddy}$ leads to an increase (decrease) in $S_1-S_2$, thereby strengthening (weakening) tropical convection and leading to an equatorward (poleward) shift of the edge of the Hadley cell, $\theta_H$.  Note, for ease of demonstration of the areal changes indicated by the red arrows, and
$S_1\left(\Delta E_{\text {eddy }}>0\right)\uparrow$ and $S_2\left(\Delta E_{\text {eddy }}>0\right)\downarrow$ in (a), and $S_1\left(\Delta E_{\text {eddy }}<0\right)\downarrow$ and $S_2\left(\Delta E_{\text {eddy }}<0\right)\uparrow$ in (b), the profiles of potential temperature are linear in latitude $\theta$.   However, although they are not monotonic, as seen in figure \ref{fig:fig03}(a), the essential behavior shown above is the same.}}
\label{fig:fig05}
\end{figure}

Our simple extension of the \citet{held1980} theory incorporates the heat transfer enhanced by synoptic eddies near the edge of the Hadley cell. 
The original theory relies on the equal area constraint of equation \eqref{eq:thetal}, rewritten here as 
\begin{align} \label{eq:constraint1}
\int_{0}^{\theta_H}\eta_Ld\theta = 0.
\end{align}
As noted above, this constraint assumes that the Hadley cell is in quasi-isolation.  However, if synoptic eddies transport heat out of the Hadley cell to mid-latitudes, we can modify the constraint as
\begin{align} \label{eq:constraint2}
-\frac{1}{\tau}\int_{0}^{\theta_H}\eta_Ld\theta=\Delta \text{E}_\text{eddy},
\end{align}
where, $\Delta \text{E}_\text{eddy}$ is the synoptic eddy driven heat flux from the Hadley cell to the mid-latitudes, including contributions from adiabatic warming associated with upper level eddy momentum flux. 

In the tropics, robust tropical convection transports an excess of heat towards higher latitudes, resulting in a negative anomaly, $\eta_L=\Theta_L-\Theta_E$, as shown in Fig.~\ref{fig:fig03}(a).  
Thus, a tropical region in a latitude range from $0$ to $\theta^*$ has positive $-\frac{1}{\tau}\eta_L$. 
Let $S_1 = -\frac{1}{\tau}\int_{0}^{\theta^*}\eta_Ld\theta$, $-S_2= -\frac{1}{\tau}\int_{\theta^*}^{\theta_H}\eta_Ld\theta$, 
and $-S_3= -\frac{1}{\tau}\int_{\theta_H}^{\pi/2}\eta_Ld\theta$, where $S_1$, $S_2$, and $S_3$ are all positive. 
Hemispheric energy conservation implies that $S_1=S_2+S_3$, which is equivalent to
an equal area constraint across the entire hemisphere:
\begin{align}  \label{eq:constraint3}
-\frac{1}{\tau}\int_{0}^{\frac{\pi}{2}}\eta_Ld\theta = 0, 
\end{align}
thereby establishing  that $S_3=\Delta\text{E}_\text{eddy}$.

The consequences of the variation of the turbulent heat transport at the edge of the Hadley cell are shown in Fig. \ref{fig:fig05}. When $\Delta\text{E}_\text{eddy}$ decreases (red arrows), $S_3$ and thus $S_1-S_2$ decrease.  This reduction is associated with the weakening of tropical convection and the associated anomalous potential temperature, and a poleward shift in the Hadley cell associated with an increase in $\theta_H$.  By parity of reasoning, the opposite scenario is realized when $\Delta\text{E}_\text{eddy}$ increases.

\section{\label{sec:conclusion} Conclusion}

The key dynamical features of the Hadley cell, including its meridional extent, the intensity of sub-tropical jets, and the tropical trade winds, are all influenced by unstable baroclinic
waves in the mid-latitudes.  The basic dynamical constraint for the Hadley circulation is angular momentum conservation. This constraint introduces a departure of the potential temperature from radiative-convective equilibrium near the poleward edge of the Hadley cell, which is used to drive large-scale atmospheric dynamics in the mid-latitudes. In particular, this departure drives baroclinic instability near the poleward edge of the Hadley cell, which results in poleward heat flux by synoptic eddies and is treated with a thermal eddy diffusivity $D$.  

We have extended the Hadley circulation model of \citet{held1980} to include the interaction between the tropics and the mid-latitudes.  Their model scaled the meridional size of the Hadley cell using the thermal Rossby number $R$.  Numerical simulations, however, reveal that mid-latitude baroclinic synoptic eddies influence the dynamical structure and meridional extent  of the Hadley cell. The simple theory presented here captures this behavior. 

A central question concerns the impact of global warming on the response of the Hadley circulation.  Our simple theory shows the importance of large-scale mid-latitude atmospheric dynamics.  Because the high-latitudes warm faster, the mean baroclinicity in the mid-latitudes is impacted non-uniformly.  In this context, recently observed changes in large-scale atmosphere dynamics  may reflect stronger interactions between the tropics  and the mid-latitudes.  This highlights the importance of having a unified theoretical framework linking the dynamics between these regions.

%%%%%%%%%%%%%%%%%%%%%%%%%%%%%%%%%%%%%%%%%%%%%%%%%%%%%%%%%%%%%%%%%%%%%
% FIGURES---INSERT NEAR IN-TEXT DISCUSSION
%%%%%%%%%%%%%%%%%%%%%%%%%%%%%%%%%%%%%%%%%%%%%%%%%%%%%%%%%%%%%%%%%%%%%
%%  Enter figures near where they are discussed within the document.
%%
%
%\begin{figure}[t]
%  \noindent\includegraphics[width=19pc,angle=0]{figure01.pdf}\\
%  \caption{Enter the caption for your figure here.  Repeat as
%  necessary for each of your figures. Figure from \protect\cite{Knutti2008}.}\label{f1}
%\end{figure}

%\clearpage
%%%%%%%%%%%%%%%%%%%%%%%%%%%%%%%%%%%%%%%%%%%%%%%%%%%%%%%%%%%%%%%%%%%%%
% ACKNOWLEDGMENTS
%%%%%%%%%%%%%%%%%%%%%%%%%%%%%%%%%%%%%%%%%%%%%%%%%%%%%%%%%%%%%%%%%%%%%
\acknowledgments
We gratefully acknowledge support from the Swedish Research Council Grant No. 638-2013- 9243.
{\color{black}WM acknowledges the Basic Science Research Program of the National Research Foundation of Korea (NRF) funded by the Ministry of Education (RS-2023-00281017) 
and the Learning \& Academic research institution for Master’s and PhD students, and Postdocs (LAMP) Program of the National Research Foundation of Korea (NRF) grant 
funded by the Ministry of Education (RS-2023-00301702)}

%%%%%%%%%%%%%%%%%%%%%%%%%%%%%%%%%%%%%%%%%%%%%%%%%%%%%%%%%%%%%%%%%%%%%
% DATA AVAILABILITY STATEMENT
%%%%%%%%%%%%%%%%%%%%%%%%%%%%%%%%%%%%%%%%%%%%%%%%%%%%%%%%%%%%%%%%%%%%%
% 
%
\datastatement
No data have been used in this publication apart from the simple solutions plotted in Figures 2 and 3.

\bibliographystyle{ametsocV6}
%\bibliography{globalcirculation}

\end{document}